\documentclass[oneside,12pt]{amsart}
\usepackage[utf8]{inputenc}
\pdfoutput=1
\usepackage{geometry}
\usepackage{lmodern}
\usepackage{amsmath}
\usepackage{booktabs}
\usepackage{placeins}
\usepackage[ruled,vlined]{algorithm2e}
\usepackage{caption}
\usepackage{comment}
\usepackage{subcaption}
\usepackage{pdflscape}
\usepackage{amssymb}
\usepackage{adjustbox}
\usepackage{amsthm}
\usepackage{thmtools,thm-restate}
\usepackage{color}
\usepackage[usenames,dvipsnames]{xcolor}
\usepackage{graphicx}
\usepackage{booktabs}
\usepackage{soul}
\usepackage[countmax]{subfloat}
\usepackage{float}
\usepackage{rotfloat}
\usepackage{setspace}
\usepackage{esint}
\usepackage[para]{threeparttable}
\usepackage[authoryear]{natbib}
\definecolor{MyDarkBlue}{rgb}{0,0.08,0.45}
\usepackage[unicode=true,pdfusetitle, bookmarks=true,bookmarksnumbered=false,bookmarksopen=false, breaklinks=false,pdfborder={0 0 1},backref=section,colorlinks=true, linkcolor = MyDarkBlue, citecolor = MyDarkBlue]{hyperref}
\usepackage{breakurl}
\usepackage{mathtools}
\usepackage{array}
\usepackage{siunitx}

\DeclarePairedDelimiter\floor{\lfloor}{\rfloor}
\newcolumntype{d}{S[
    input-open-uncertainty=,
    input-close-uncertainty=,
    parse-numbers = false,
    table-align-text-pre=false,
    table-align-text-post=false
 ]}

\usepackage{pstricks}
\usepackage{tikz}
\usetikzlibrary{arrows}
\usetikzlibrary{shapes}
\usetikzlibrary{plotmarks}

\PassOptionsToPackage{normalem}{ulem}
\usepackage{ulem}
\allowdisplaybreaks

\makeatletter

\numberwithin{equation}{section}

\newtheorem{proposition}{Proposition}

\newcommand{\downgeq}{\mathbin{\text{\rotatebox[origin=c]{90}{$\geq$}}}}
\newcommand{\upgeq}{\mathbin{\text{\rotatebox[origin=c]{90}{$\leq$}}}}
\newcommand{\downg}{\mathbin{\text{\rotatebox[origin=c]{90}{$>$}}}}
\newcommand{\upg}{\mathbin{\text{\rotatebox[origin=c]{90}{$<$}}}}

\def\eproof{\hbox{\hskip3pt\vrule width4pt height8pt depth1.5pt}}

\title{Multiplexing in Networks and Diffusion}

\author{Arun G. Chandrasekhar$^{\ddagger}$}
\author{Vasu Chaudhary$^{\S}$}
\author{Benjamin Golub$^{\S}$ }
\author{Matthew O. Jackson$^{\dagger}$ }

\date{This Draft: \today}
\thanks{$^{\ddagger}$Department of Economics, Stanford University; NBER; J-PAL}
\thanks{$^{\dagger}$Department of Economics, Stanford University; Santa Fe Institute}
\thanks{$^{\S}$Department of Economics, Northwestern University}
\thanks{Yann Calvó López provided exceptional research assistance. We thank Paul Goldsmith-Pinkham for helpful conversations.  Research was supported by NSF grants SES-1629328 and SES-2018554.  Chandrasekhar acknowledges the Alfred P. Sloan Foundation for financial support.}

\begin{document}

\begin{abstract} 

Social and economic networks are often multiplexed, meaning that people are connected by different types of relationships---such as borrowing goods and giving advice. We make two contributions to the study of multiplexing and the understanding of simple versus complex contagion. On the theoretical side, we introduce a model and theoretical results about diffusion in multiplex networks. We show that multiplexing impedes the spread of simple contagions, such as diseases or basic information that only require one interaction to transmit an infection.  We show, however that multiplexing enhances the spread of a complex contagion when infection rates are low, but then impedes complex contagion if infection rates become high. 
On the empirical side, we document empirical multiplexing patterns in Indian village data.  We show that relationships such as socializing, advising, helping, and lending  are correlated but distinct, while commonly used proxies for networks based on ethnicity and geography are nearly uncorrelated with actual relationships. 
We also show that these layers and their overlap affect information diffusion in a field experiment. The advice network is the best predictor of diffusion, but combining layers improves predictions further. Villages with greater overlap between layers---more multiplexing---experience less overall diffusion. Finally, we identify differences in multiplexing by gender and connectedness. These have implications for inequality in diffusion-mediated outcomes such as access to information and adherence to norms.


\end{abstract}

\maketitle

\thispagestyle{empty}

\setcounter{page}{0} \newpage

\section{Introduction}

People maintain many different types of relationships.
For example, college students' partners in social activities overlap with, but also differ from, the people to whom they turn during times of stress or for academic collaboration  \citep{morelli2017empathy,jacksonnsy2024}.
Villagers in south India seek advice from one set of people, borrow money from a different set, and borrow kerosene and rice from yet a third set, and these sets partly overlap.  Moreover,
these networks have distinct properties (e.g., link frequencies, clustering coefficients, distributions of centrality, etc.) \citep{banerjeecdj2013}. 
Importantly, even though these different network ``layers'' all serve different purposes, people talk with each other in each dimension and can spread disease across all layers. 
How does the fact that people are embedded in a number of different, overlapping networks with distinct properties affect the diffusion of information or other things that can spread via any contact?

The coexistence of distinct types of relationships among the same population is known as \emph{multiplexing} (see, e.g., \cite{kivelaetal2014}), and the interdependence of different types of relationships has been discussed since  \cite{simmel1908}.
Although there are case studies of multiplexed relationships, many basic questions remain open concerning the  patterns of multiplexing in social and economic relationships, as well as how multiplexing impacts diffusion.

As we show here, multiplexing turns out to be a key to understanding differences between `simple' and `complex' contagion.  Simple contagion refers to situations where just one interaction with an infected individual is enough to transmit a disease, belief or behavior,  while complex contagion refers to situations where multiple interactions/reinforcement is needed to transmit a belief or behavior \citep{granovetter1978,centola2018,beaman2021can}.
Previous research has shown that local clustering patterns in networks are important in whether a complex contagion can be initiated \citep{centola2010}.  As we prove here, multiplexing determines the overall spread of a complex contagion, and in ways that
can differ from how it affects the spread of a simple contagion.    

In this paper, we make two main contributions, one theoretical and one empirical, about patterns of multiplexing and how it impacts diffusion. 

Our theoretical contribution is the development of a model of, and theoretical results on, how multiplexing determines diffusion.
We first model how multiplexing affects a simple diffusion or contagion process in which a person may be infected/informed by any single infected other (called \emph{simple} contagion).  
We introduce a definition capturing what it means for an individual to be more multiplexed in one multilayered network than another. 
We prove that a more multiplexed individual is less likely to become infected for any given probability of neighbor infection.   Building on this result, we demonstrate that in a standard SIS (Susceptible-Infected-Susceptible) contagion model, the steady-state infection rate decreases as individuals become more multiplexed as long as transmission is not too negatively correlated across layers. These results can be summarized by saying that multiplexing impedes ``simple'' diffusions. 
We then develop a theory of how multiplexing impacts ``\emph{complex}'' diffusion processes---in which people only become infected or adopt a new behavior/practice if they experience sufficiently many interactions with infected alters. 
We show that multiplexing enhances diffusion when the infection rate is low, but then impedes it when the infection rate is high.  The nonmonotonicities identified by our theory reveal that multiplexing has subtle implications for complex contagion.

The intuition behind the contrast between simple and complex contagion is as follows. In simple diffusion a single infected interaction suffices to infect a node. Diversifying contacts across multiple individuals increases the probability that at least one contact is infected, compared to concentrating multiple interactions with the same person. Consequently, multiplexing impedes simple contagion, provided that transmission is not too negatively correlated across layers.  Complex diffusion involves more subtle tradeoffs. Consider an individual with two links in different layers. Whether infection is more likely when these links connect to the same friend or to different friends depends on the infection's prevalence. When infection is already widespread, the individual likely needs only one additional interaction to reach the contagion threshold. In this regime, diversifying links increases the chance of contacting an infected individual, paralleling our discussion simple diffusion, and thus less multiplexing enhances infection. Conversely, when infection rates are low, the individual more likely requires multiple additional contacts to become infected. The probability of transmission occurring on both links is higher when they connect to the same person, since only that individual need be infected, rather than requiring both of two separate contacts to be infected. Therefore, when overall infection rates are low (less contagious processes), greater multiplexing accelerates complex diffusion.

Before presenting the theoretical results, we present our other main contribution which is to document patterns of multiplexing, as well as how it impacts a diffusion empirically.  We show that multiplexing varies significantly across settings, that multiple layers are differently useful in predicting a diffusion, and that it impedes diffusion in an empirical application that matches up with a simple diffusion.  
In particular, our empirical analysis comes in three parts.

First, using two large datasets of multiplexed networks we examine the correlation across network layers. We document both significant correlation between different network layers (types of relationships) and meaningful differences in their patterns. 
We show that layers of informational relationships, financial relationships, and social relationships, among others, exhibit strong correlations in a sample of 143 villages in Karnataka, India, comprising nearly 30,000 households 
\citep{banerjeecdj2013,banerjeecdj2019,banerjee2018changes}. At the same time, different layers display distinct patterns and differ in density and other network statistics. 
We also show that proxies for social relationships that are commonly used in the peer effects literature, such as geographic proximity or co-ethnicity (in our data, being members of the same \emph{jati}, or subcaste), are nearly orthogonal to the other layers. This suggests that relational variables constructed based on geographic or ethnic covariates can fail to serve as accurate surrogates for actual social and economic relationships.

Second, we show that distinctions among layers are substantively important for the study of economic outcomes, specifically the diffusion of information and behaviors. Even though people have contacts and can influence each other in any network layer, we find that the different layers are differently predictive of diffusion and combine to form a nuanced overall picture.
Using data from a randomized controlled trial of information diffusion, 
we show that some layers are more predictive of diffusion than others---with an ``advice'' layer standing out---and moreover that using a combination of layers yields predictions significantly better than those based on any single layer. 
A combination of layers also affords better predictions than using the union or intersection of layers. These findings indicate that the layers are not simply noisy observations of a latent one-dimensional network, but instead contain information richer than any one-dimensional summary. Without properly accounting for the multiplexed nature of relationships, researchers can arrive at misleading conclusions about peer effects and influence.\footnote{An additional finding shows that links are most usefully viewed as multi-dimensional. We show that, although the jati layer is the least predictive of diffusion and not a good proxy for actual relationships, combining it with other layers significantly improves diffusion predictions. Thus, although jati is not a good substitute for elicited network data, it is a valuable complement. As we discuss below, using the jati variable drastically over-predicts links within jati, and under-predicts them across jati.  One conjecture as to why the jati layer helps in predicting diffusion is that patterns of information passing on the network are related to jati.}  

Third, as another important empirical observation that helps motivate our theoretical investigation: villages that are ``more multiplexed'' (have more strongly correlated layers) experience less diffusion.

We close the paper 
with observations about how multiplexing varies with individual characteristics and some implications for issues of inequality.  We find that women's networks display significantly more multiplexing than men's networks, and that multiplexing correlates negatively with the number of connections a person has. Given our theoretical and empirical evidence showing how multiplexing can impede simple diffusions, this suggests that multiplexing could function as a channel limiting women's exposure to information. More broadly, demographic differences in multiplexing imply that network-mediated contagions work differently in different subpopulations---a rich topic that we believe deserves further study. 

\

Our findings are orthogonal and complementary to the research that has examined how clustering and homophily patterns affect simple and complex contagion \citep{granovetter1978,morris2000,jacksony2005,centola2010,golubj2012,aral2012identifying,jacksons2017,mosleh2021shared}. 
Those focus on local densities of links and community patterns that are necessary for a behavior or infection to spread from one part of a network to another, while the multiplexing examines differences in layers and correlations between those layers and how those affect overall levels of spreading.  These are distinct both in the questions asked, and the reasoning and intuitions that the answers provide.

The literature on multiplex networks has begun to grow in the last decade \citep{contractor2011network,boccaletti2014structure,kivelaetal2014,dickison2016multilayer,bianconi2018multilayer}. The recognition that people are involved in different types of relationships dates to some of the original works on network analysis (e.g., \cite{simmel1908}), and instances of the fact that different layers can serve different roles have been analyzed over time \citep{wassermanf1994, becker2020multiplex}.
More recent studies have shown that distinguishing between different networks and tracking their interplay can be important in understanding cooperative behavior \citep{atkissonetal2019,cheng2021theory} as well as understanding play in network games and targeting policies to influence it \citep{walsh2019games,zenou2024games}. New (independent) work by \cite{shi2025multiplex} also finds that network layers are differently predictive of diffusion of a behavior in a rural setting, and proposes a method of decomposing the impact of layers. 

Our contributions to the literature on multiplex networks are threefold.
First, we provide some of the first detailed statistical analyses of how multiple layers relate to each other in empirical social networks.
Second, we show how different layers---as well as the level of correlation between layers---predict diffusion outcomes.
This suggests that unidimensional theories of diffusion and contagion can miss important factors
that determine the extent of diffusion.
Third, we introduce a model and develop a new theoretical analysis of how correlation between layers impacts diffusion, which provides a basis for interpreting our empirical observations about multiplexing and diffusion.  

While some theoretical work has examined simple \citep{hu2013percolation,larson2023risk}
and complex 
\citep{yaugan2012analysis,zhu2019social,kobayashi2023dynamics} diffusion on multiplexed networks,
previous analyses have focused on independently distributed layers.  Such diffusion models are a more direct extension of diffusion on one layer and the proofs in the existing literature leverage that fact.   Our analysis examines how changes in layer overlap affect diffusion.  In addition---and in contrast to prior models---our model also allows for interactions (such as conversations or information transmissions) to be correlated across layers, even conditional on links. 

Our findings on the impact of multiplexing on diffusion can help inform a nascent and important literature on the incentives to form multiplexed networks
\citep{billand2023model,san2024multiplexed}.
For example, a series of empirical studies on rural developing economies emphasize the role social networks play in risk-sharing arrangements \citep{townsend1994,fafchamps2007risk,ambrus2014consumption,cai2015social}.   
This raises a fundamental question: If individuals primarily organize their relationships around risk-sharing and multiplex other relationships on top of the risk-sharing relationships, how might these structures affect the diffusion of new information or technologies?  Both our empirical findings and theoretical results shed light on this issue.


\section{The Structure of Multiplexed Networks}

\subsection{Two Datasets of Multiplexed Networks}

We study two different data sets of multiplex networks in a total of 143 villages, both from the state of Karnataka, India, covering a total population of nearly 30,000 households. 
\\
\paragraph{The Microfinance Network Data}

The first dataset, which we refer to as the \emph{microfinance village sample}, includes 75 villages surveyed in 2012 \citep{banerjeecdj2013,banerjee2018changes}.
That study obtained a complete census of the 16,476 households across these 75 villages. From 89.14\% of these households\footnote{Given our focus on undirected graphs, we elicit a link as long as at least one of the two households on either end is sampled. With 89.14\% of the households being sampled, for  two arbitrary nodes $i$ and $j$, we compute P($i$ or $j$ in  sample) = $1 - (1-0.89)^2 = 0.9879$.} the researchers also collected socio-economic network data
on seven types of ties:
\begin{itemize}
\item[(1)] social: to whose home does the respondent go and who comes to their home, as well as which close relatives live outside their household;
\item[(2)] kerorice: from whom does the respondent borrow kerosene/rice and to whom does the respondent lend these goods; 
\item[(3)] advice: to whom does the respondent give information/advice;
\item[(4)] decision help: to whom does the respondent turn for help with an important decision;
\item[(5)] money: if the respondent  suddenly needed to borrow 50 rupees for a day, to whom would they turn, and who would come to them with such a request;
\item[(6)] temple: if the respondent goes to a temple, church, or mosque, who might accompany them;
\item[(7)] medic:  if the respondent had a medical emergency alone at home, whom would they ask for help in getting to a hospital.
\end{itemize}

Additionally, we have information on the {jati} (subcaste) and GPS coordinates for each household. This allows us to construct jati networks (in which pairs from the same subcaste are linked) and geographic networks, whose edges are labeled by distance in physical space. Variables of these types have been used as proxies for social networks in prior studies \citep[e.g.,][]{sacerdote2001peer,fafchamps2007risk,munshir2009}.
\\
\paragraph{The Diffusion RCT Network Data}

The second dataset, which we refer to as the \emph{RCT village sample}, includes data on 68 villages collected from a separate diffusion experiment by \cite{banerjeecdj2019}.

The network data was collected in a manner similar to that of the Microfinance Village Sample. The surveys elicited information about the following layers:
\begin{itemize}
\item[(1)] social: to whose home does the respondent go and who comes to their home to socialize; 
\item[(2)] kerorice: from whom does the respondent borrow kerosene/rice or small amounts of money and to whom does the respondent lend these goods; 
\item[(3)] advice: to whom does the respondent give information/advice;
\item[(4)] decision: to whom does the respondent turn for help with an important decision?
\end{itemize}
While we have jati information for the RCT villages, we lack GPS data for this sample.
\\

\paragraph{Network Construction and Notation}

A link is present in a given layer if either household named the other household in one of the questions in that category (e.g., we code a kerorice link if either household reports borrowing kerosene or rice from or lending it to the other household).  In terms of notation, we define a multi-layered, undirected network for each village $v$,
for layer $\ell = 1,\ldots , L$, with $g_{ij,v}^\ell = 1$ if either household $i$ or $j$ reported having a relationship of type $\ell$. We add another layer where $i$ and $j$ are linked if they belong to the same jati. For the Microfinance Village Sample, where GPS data are available, we construct a weighted graph where the $ij$ entry is the geographic distance between the two households. 

The \emph{union} layer has a link present if a link exists in any layer. The \emph{intersection} layer has a link present if it exists in all layers.\footnote{Both of these definitions {include} the jati layer but {exclude} geography, since we are able to define the geographic layer for only one of our datasets, and it is a weighted network in any case. We make these definitions to maintain consistency of the meaning of the union and intersection layers across the two data sets.}

We also build a weighted and directed network whose edge weights are the sums of indicators for links in all directed layers (using the raw directed nomination matrices, thus excluding jati and geography).  We call this the \emph{total} network.  Finally, below we describe another weighted and directed aggregate network that we build from the principal component analysis.

\subsection{The Diffusion Randomized Controlled Trial}

In the second dataset we also have the data from a randomized controlled trial (RCT) studying diffusion in these villages, which is the subject of \cite{banerjeecdj2019}. This RCT provides cleanly identified estimates of diffusion, allowing us to examine how diffusion varies with different aspects of multiplexed networks. 
Specifically, in each village either 3 or 5 individuals (determined uniformly at random) were seeded with information about a promotion.  Villagers could obtain a non-rivalrous chance to win either cash prizes or a mobile phone by calling in to register for the promotion.

Thus, the experiment induced the diffusion of a non-rivalrous, valuable piece of information. The outcome variable of interest is the number of households that registered.  There was exogenously randomized variation in the position of the random seeds in the network, and more central seeds caused larger diffusions. We use our data on multiplexing to examine how diffusion depends on the network statistics of the seeds in various network layers, used individually and in combination, and on network multiplexing levels.

Additional details on the two datasets appear in Appendix Section \ref{appdata}.

\subsection{Correlations across Network Layers}
\

Across both our samples, social, informational, and financial relationships often overlap---but the alignment across layers is far from complete. The social layer is denser than the other layers and has the highest level of clustering, while the decision layer exhibits higher variance of node degree than other comparable layers (e.g. advice). Jati based links are strikingly different: they form a much denser network than other layers, but serve as a poor proxy for other types of relationships---too dense, too clustered, and too homophilous to predict the other layers. Detailed descriptive statistics can be found in Supplementary Table \ref{table:desc}. 

Fig.~\ref{fig:corr_heatmap} reveals three clear patterns as we examine pairwise correlations between layers. 

 \begin{figure}[h]
	\centering
	\subfloat[Microfinance Villages]{%
		\includegraphics[width=0.9\linewidth]{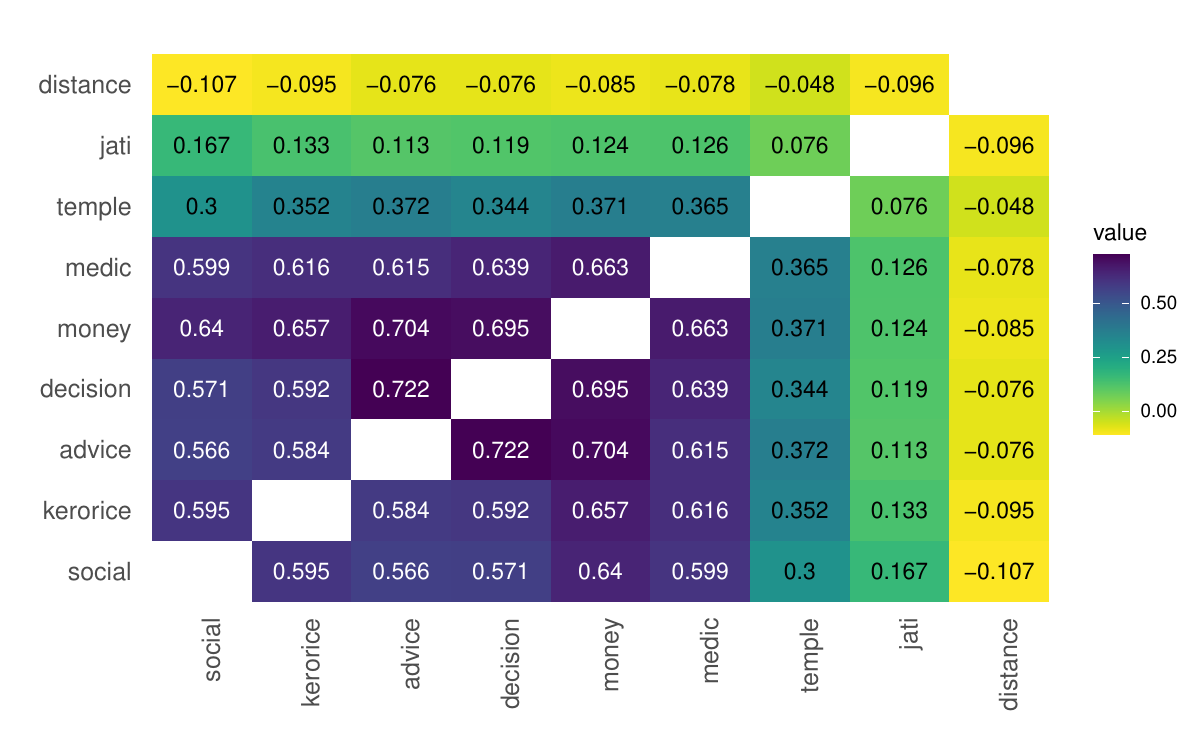}
	}
	
	\subfloat[RCT Villages]{%
		\includegraphics[width=0.9\linewidth]{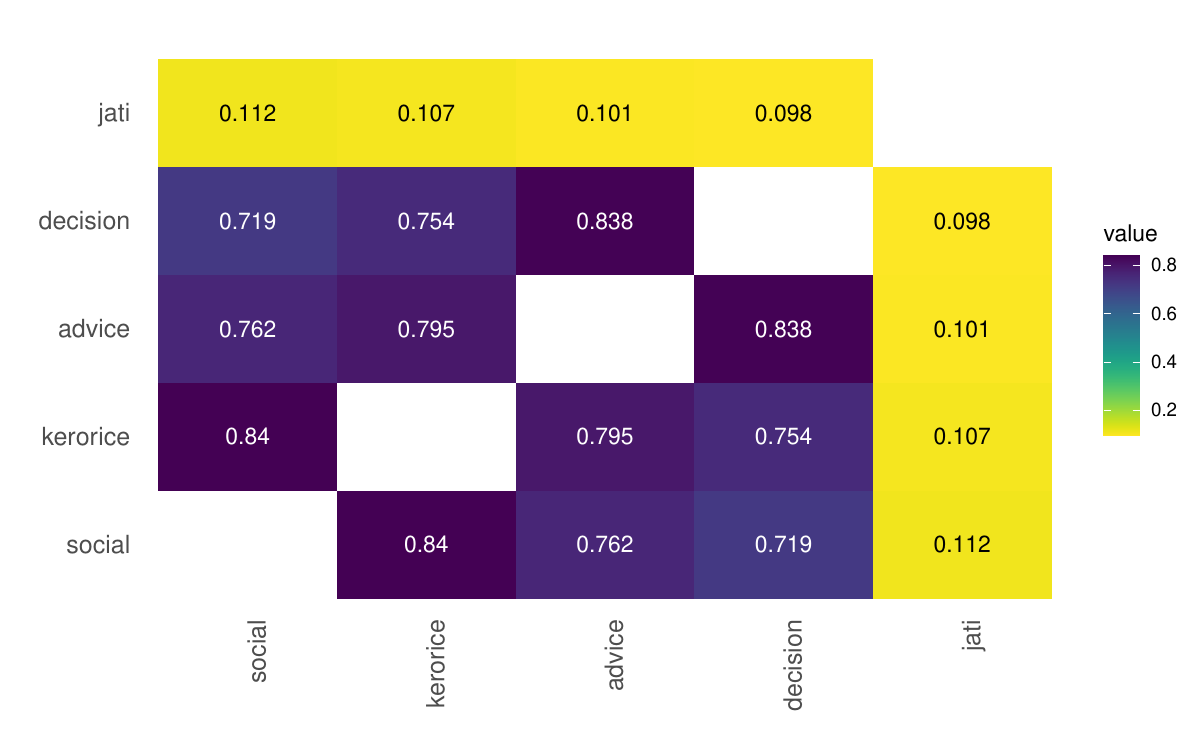}
	}
	
	\caption{Correlation Heatmaps}
	\label{fig:corr_heatmap}
\end{figure}

First, there are consistently high correlations between layers in both data sets---above 0.5 for most layer pairs.  Second, the exceptions are the distance, jati, and temple layers.  The jati and distance layers are almost uncorrelated with the other layers,\footnote{This does not mean, for instance, that there is not substantial jati-based homophily in these data.  The low correlation comes from the fact that the jati layer dramatically over-predicts relationships compared to other layers, so it has many 1's where there are 0's in the other layers.}$^,$\footnote{Distance is higher when people live far from each other and are thus less likely to be linked, all else held equal; this explains the negative signs.} while the temple layer has an intermediate level of correlation with others.
Third, the layers are more highly correlated in the RCT villages compared to the microfinance villages.

\subsection{Principal Components and the Backbone Network}
\

To further analyze the network structure across layers, we apply a principal component analysis (PCA) to the multigraph of each village. 
We perform the analysis with all of the layers (excluding the synthetic \emph{union} and \emph{intersection} layers).
We treat each pair of households (in a given village) as an observation, yielding $\sum_v \binom{n_v}{2}$ observations, where $n_v$ is the number of households in village $v$, and the number of dimensions is the number of layers $L$ in the given sample. 
(Details for how the PCA was implemented can be found in Appendix Section \ref{app:pca}.) 

\begin{figure}[h]
	\centering
	\subfloat[Principal Components: Microfinance Villages]{%
		\includegraphics[width=0.45\linewidth]{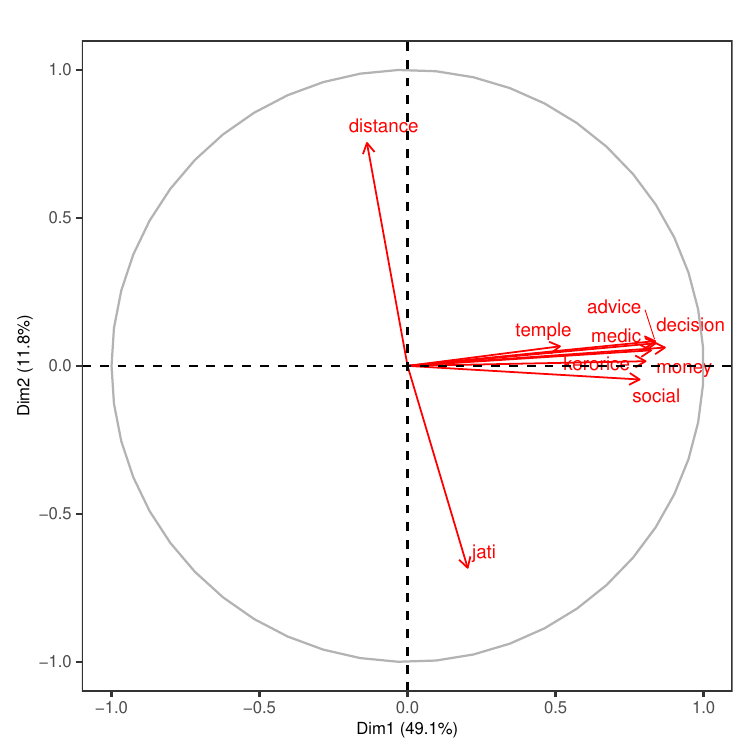}
	}
	\hfill
	\subfloat[Principal Components: Diffusion RCT]{%
		\includegraphics[width=0.45\linewidth]{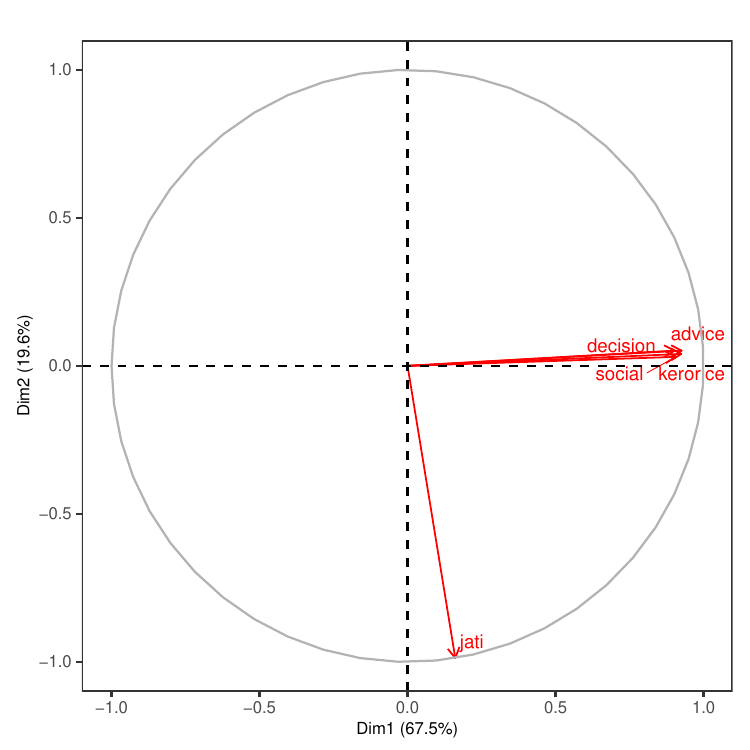}
	}
	\caption{Principal Component Analysis with All Layers}
	\label{fig:pca}
\end{figure}

The first principal component aligns closely with most relationship layers and captures $49$\% of variation in the microfinance sample and $72$\% in the RCT sample (Fig \ref{fig:pca} panels A and B). Jati, by contrast, loads primarily on the second component, consistent with its limited correlation with other ties. Geographic distance is negatively aligned with jati, reflecting residential clustering by caste. Complete loadings for each layer appear in Supplementary Tables \ref{tab:loadings_mf} and \ref{tab:loadings_rfe}.

\begin{figure}[h]
    \centering
    \begin{subfigure}[t]{0.48\linewidth}
        \includegraphics[width=\linewidth]{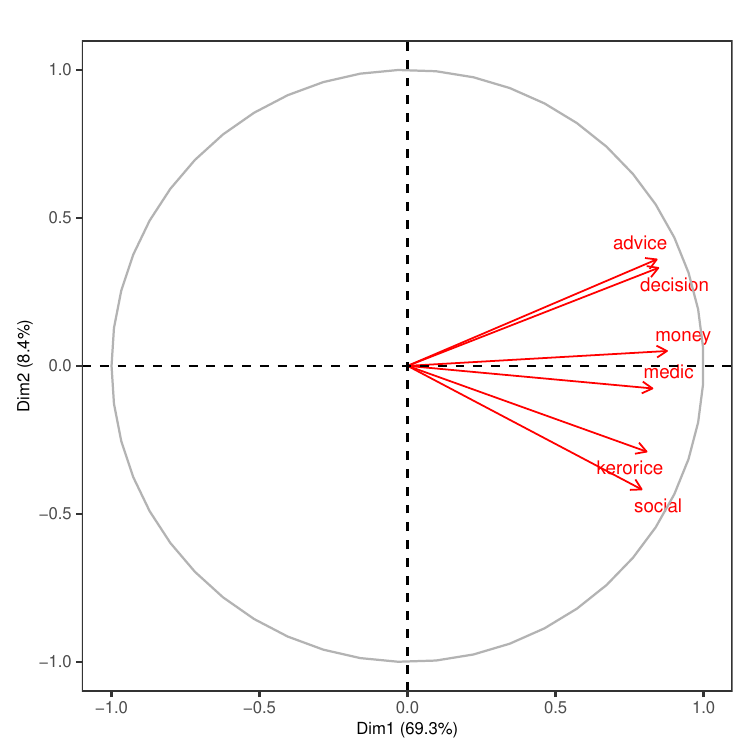}
        \caption{Principal Components: Microfinance Villages}
    \end{subfigure}
    \hfill
    \begin{subfigure}[t]{0.48\linewidth}
        \includegraphics[width=\linewidth]{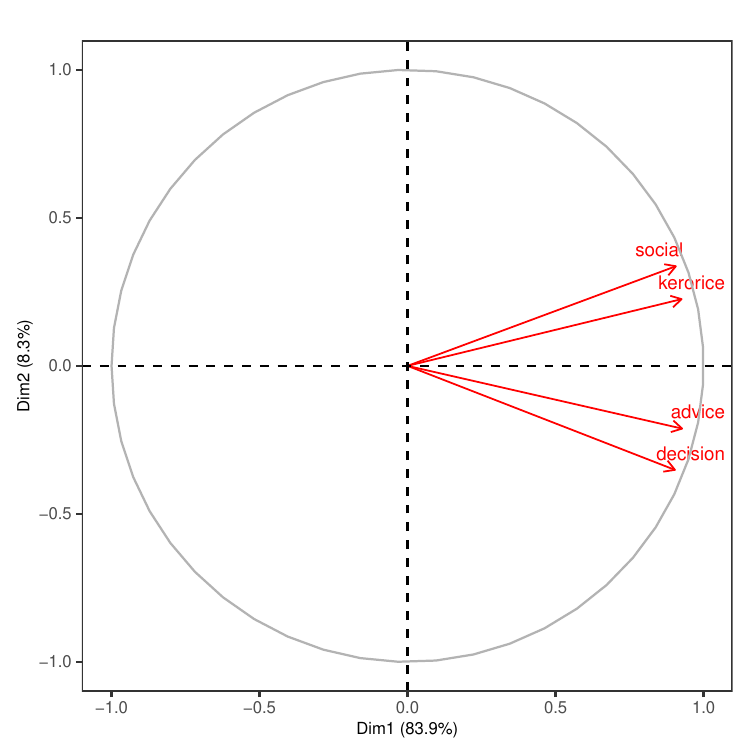}
        \caption{Principal Components: Diffusion RCT}
    \end{subfigure}
    \caption{Principal Component Analysis Excluding Jati, Geography, and Temple}
    \label{fig:pca_2}
\end{figure}

Next, in Figure \ref{fig:pca_2}  we repeat the analysis after removing the least correlated dimensions: jati, geography, and temple.\footnote{
We also redo the analysis just dropping jati and geography and keeping temple in Supplemental Appendix Figure \ref{fig:pca_withtemple}. Temple is sparse and essentially orthogonal to the other dimensions.  
}  This allows us to zoom in on the correlation patterns among the social and economic layers.

Figure \ref{fig:pca_2} displays the relationship between the layers where we again project them on the first two principal components. In Panel A, we can see three distinct groupings of similar layers in the microfinance villages (advice-decision, money-medic, and kerorice-social). In Panel B, there appear to be two distinct groupings in the RCT villages (advice-decision, kerorice-social), with the first component now explaining 70\% and 83\% of the variance across the two samples, respectively.

To formally capture the correlation structure of these network layers, we use the principal components to construct a continuous weighted network, which we call the \emph{backbone}. The backbone network is built using the first $K$ principal components, where the optimal $K$ is derived from the so-called ladle plot from \cite{luoli2016laddle} (see Appendix Section \ref{app:pca} for details). 
The backbone provides a low-dimensional index which reduces the multiplex data to a synthetic structure by projecting multidimensional links onto the top principal components.


\section{Multiplexing and Information Diffusion}\label{sec:determinant_diffusion}

The empirical analysis demonstrates that multiplexed networks in our data are rich and embed important information that would be lost by collapsing them into a single summary measure.  A natural next question is 
how this distinction between layers matters for outcomes of interest. Here we focus on diffusion in the RCT villages.

We proceed as follows.  Based on prior work, we would expect that more central seeds should lead to greater diffusion \citep{banerjeecdj2013,banerjeecdj2019}.  Those papers defined a single network by using the union network and computed diffusion centrality based on that.
However, given our rich multiplex data, we note that the seeds' diffusion centrality differs across layers.   
Thus, we examine which layer is the most predictive of diffusion in the RCT if we were to compute diffusion centrality based on that layer alone.

We use a specific diffusion centrality measure developed in \cite{banerjeecdj2013} and further studied in \cite{banerjeecdj2019}. 
The details of that measure appear in Appendix Section \ref{app:var_def}.  

We then can calculate how diffusion varies with the diffusion centrality of the randomly assigned seed set under layer $\ell$ by regressing
\begin{align}\label{eq:ols-1}
y_v = \alpha^\ell + \beta^\ell \cdot DC_v^\ell +  X_v \Gamma^\ell + \epsilon_{v,\ell}
\end{align}
where   $y_v$ represents the number of calls received from village $v$ (a measure of diffusion of information), and $X_v$ includes controls for number of households, its second and third powers, and number of seeds assigned in that village. We standardize all the regressors.

Table \ref{tab:table_dcseed} depicts how differently the layers predict diffusion based on our specification in (\ref{eq:ols-1}). (Supplementary Appendix Figure \ref{fig:ols} plots the $90$\% and $95$\% confidence intervals.)

\begin{table}
\centering
\caption{Seed Set Diffusion Centrality}
\label{tab:table_dcseed}
\scalebox{0.7}{
\begin{threeparttable}
\centering
\begin{tabular}[t]{lccccccccc}
\toprule
\multicolumn{1}{c}{ } & \multicolumn{9}{c}{No. Calls Received} \\
\cmidrule(l{3pt}r{3pt}){2-10}
  & 1 & 2 & 3 & 4 & 5 & 6 & 7 & 8 & 9\\
\midrule
Social & \num{4.266} &  &  &  &  &  &  &  & \\
 & (\num{1.820}) &  &  &  &  &  &  &  & \\
 & {}[\num{0.022}] &  &  &  &  &  &  &  & \\
Kero/Rice &  & \num{5.466} &  &  &  &  &  &  & \\
 &  & (\num{2.326}) &  &  &  &  &  &  & \\
 &  & {}[\num{0.022}] &  &  &  &  &  &  & \\
Advice &  &  & \num{6.410} &  &  &  &  &  & \\
 &  &  & (\num{2.416}) &  &  &  &  &  & \\
 &  &  & {}[\num{0.010}] &  &  &  &  &  & \\
Decision &  &  &  & \num{3.137} &  &  &  &  & \\
 &  &  &  & (\num{2.226}) &  &  &  &  & \\
 &  &  &  & {}[\num{0.164}] &  &  &  &  & \\
Jati &  &  &  &  & \num{1.161} &  &  &  & \\
 &  &  &  &  & (\num{1.559}) &  &  &  & \\
 &  &  &  &  & {}[\num{0.459}] &  &  &  & \\
Union &  &  &  &  &  & \num{1.110} &  &  & \\
 &  &  &  &  &  & (\num{1.752}) &  &  & \\
 &  &  &  &  &  & {}[\num{0.529}] &  &  & \\
Intersection &  &  &  &  &  &  & \num{2.220} &  & \\
 &  &  &  &  &  &  & (\num{2.200}) &  & \\
 &  &  &  &  &  &  & {}[\num{0.317}] &  & \\
Backbone &  &  &  &  &  &  &  & \num{1.752} & \\
 &  &  &  &  &  &  &  & (\num{2.123}) & \\
 &  &  &  &  &  &  &  & {}[\num{0.412}] & \\
Total Links &  &  &  &  &  &  &  &  & \num{2.158}\\
 &  &  &  &  &  &  &  &  & (\num{1.453})\\
 &  &  &  &  &  &  &  &  & {}[\num{0.143}]\\
\midrule
Num.Obs. & \num{68} & \num{68} & \num{68} & \num{68} & \num{68} & \num{68} & \num{68} & \num{68} & \num{68}\\
R2 & \num{0.194} & \num{0.254} & \num{0.313} & \num{0.161} & \num{0.110} & \num{0.110} & \num{0.139} & \num{0.119} & \num{0.131}\\
Dep Var mean & \num{8.691} & \num{8.691} & \num{8.691} & \num{8.691} & \num{8.691} & \num{8.691} & \num{8.691} & \num{8.691} & \num{8.691}\\
\bottomrule
\end{tabular}
\begin{tablenotes}
  \small
  \item \textit{Note:} Robust standard errors are given in parentheses and p-values in square brackets. Controls added: number of households, its powers, and a dummy for number of seeds in the village. Exogenous variables are the sum of Diffusion Centrality for seeds in each village for the layer. Exogenous variables have been standardized. The total links network is the raw sum of all directed network layers (excluding jati network).
  \end{tablenotes}
  \end{threeparttable}
}
\end{table}

The advice layer stands out as the most predictive, and we see that the kerorice and social layers are also significantly predictive.   
Notably, consistent with what we observed in the correlations and principal component analysis, jati explains the least of the variation and is not significant.

Interestingly, the four synthetic networks we have mentioned that aggregate the layers in specific ways---\emph{union, intersection, total}, and \emph{backbone}---all perform worse than the individual layers with the exception of jati.  However, this appears to be rooted in the inclusion of jati in those aggregates.  In Supplement \ref{supp:tables} we recreate Table \ref{tab:table_dcseed} with aggregate layers that omit jati in their construction (this applies only to \emph{union}, \emph{intersection}, and \emph{backbone}).  This improves their performance, with the backbone network now yielding an $R^2$ second only to the advice layer.  

Given how correlated the layers are, we also 
perform a  LASSO ($\ell_1$-penalized) regression to select a sparse set of relevant variables that explain diffusion. We then use post-LASSO least squares to estimate how seed set centrality under the selected layer(s) affects diffusion. 

The regression of interest is given by
\begin{align}\label{eq:lasso}
y_v = \alpha + \sum_\ell \beta^\ell \cdot DC_v^\ell +  X_v \Gamma + \epsilon_{v}
\end{align}
where the variables are as described in \eqref{eq:ols-1} and instead of running a separate regression for each layer,
we now include all the layer variables simultaneously.  We are interested in which $\beta^\ell$ are estimated to be non-zero and the consistent estimates of these parameters.

A complication we face here is that in order to be consistent, LASSO requires a condition called irrepresentability, which requires the regressors of interest not to be excessively correlated \citep*{zhao2006model}. In our setting,  this requirement fails since the network layers are highly correlated.  To overcome this problem, we use the Puffer transformation developed by \cite{rohe2015preconditioning} and  \cite{jia2015preconditioning}, which recovers irrepresentability when the number of observations exceeds the number of variables.
See Appendix section \ref{app:lasso} for details.

\begin{figure}
    \centering
    \includegraphics[scale = 0.6]{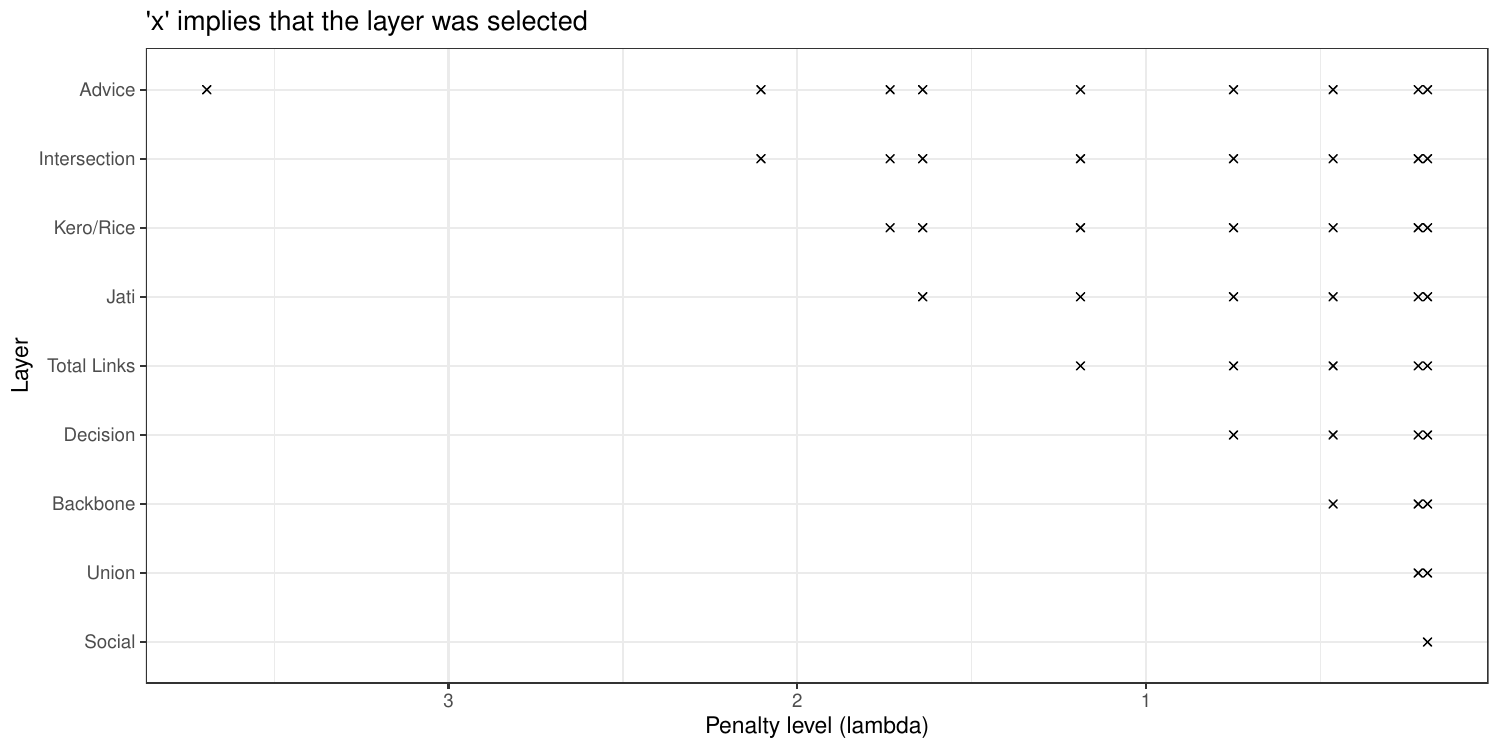}
    \caption{Lasso Selection of Layers in Predicting Diffusion}
    \label{fig:lasso}
\end{figure}

We see the results in Figure \ref{fig:lasso}, where we plot which layers are selected by the LASSO as we increase the penalty level, forcing LASSO to select fewer variables. We find that, at the highest penalty level, only the advice network layer is selected, with the post-puffer LASSO OLS regression in Table \ref{tab:postlasso} depicting a 64\% increase in diffusion relative to the mean ($ p= 0.011$). Despite the fact that multiple layers are useful in explaining 
diffusion, neither the backbone,  the union, nor the intersection network proved to be the most useful.

\begin{table}[H]
\centering
\caption{Post Puffer Lasso OLS: Seed Set Diffusion Centrality}
\label{tab:postlasso}
\centering
\begin{tabular}[t]{lc}
\toprule
\multicolumn{1}{c}{ } & \multicolumn{1}{c}{No. Calls Received} \\
\cmidrule(l{3pt}r{3pt}){2-2}
\midrule
Advice & \num{5.564}\\
 & (\num{2.117})\\
 & {}[\num{0.011}]\\
\midrule
Num.Obs. & \num{68}\\
R2 & \num{0.233}\\
Dep Var mean & \num{8.691}\\
\bottomrule
\end{tabular}
\end{table}

The fact that centrality in the advice layer is singled out as the best predictor of diffusion under sufficiently high penalty does not mean that the other layers have no impact on diffusion.  
In fact, a combination of the layers still provides significantly more
prediction than just the advice layer, as shown in Table \ref{tab:ftest1}.

\begin{table}[H]
\caption{F-test for the layers}
\label{tab:ftest1}
\centering
\begin{threeparttable}
\begin{tabular}[t]{lrrrrrr}
\toprule
layer & df & R.sq. & F-stat & p-val & F-stat marginal & p-val marginal\\
\midrule
Advice & 1 & 0.233 & 20.057 & 0.000 &  & \\
Intersection & 2 & 0.276 & 3.888 & 0.053 & 3.888 & 0.053\\
Kero/Rice & 3 & 0.281 & 2.134 & 0.127 & 0.415 & 0.522\\
Jati & 4 & 0.325 & 2.844 & 0.045 & 4.059 & 0.048\\
Total Links & 5 & 0.343 & 2.602 & 0.044 & 1.771 & 0.188\\
Decision & 6 & 0.348 & 2.159 & 0.070 & 0.478 & 0.492\\
Backbone & 7 & 0.353 & 1.851 & 0.104 & 0.416 & 0.521\\
Union & 8 & 0.353 & 1.564 & 0.164 & 0.021 & 0.884\\
Social & 9 & 0.353 & 1.349 & 0.238 & 0.026 & 0.873\\
\bottomrule
\end{tabular}
\begin{tablenotes}
\small
\item \textit{Note}: We present both cumulative and marginal F-tests as layers are successively added in the order selected by LASSO. The ``F-stat'' and ``p-val'' columns correspond to the cumulative test comparing each specification with the intercept only benchmark. ``F-stat marginal'' and ``p-val marginal'' columns correspond to the marginal test when adding a given layer.
\end{tablenotes}
\end{threeparttable}
\end{table}

Table \ref{tab:ftest1} presents both cumulative and marginal F-tests as variables are added in the order selected by LASSO. We can see that adding intersection is marginally significant above advice, and further including both kerorice and jati together yields a more complete model, with an improvement significant at the 5 percent level.\footnote{
In Appendix Table \ref{tab:ftest_short} we exclude the extra layers of intersection, union, total links and backbone, which are ``constructed'' layers that are derived from these basic layers.  F-tests include the basic layers in the order selected by the Lasso.}
Thus, even though jati serves as a poor substitute for other layers, it turns out to be a useful complement to them in predicting diffusion.

\subsection{How the Level of Multiplexing Affects Diffusion}
\label{defmultiplex}

\

Next, we examine how diffusion depends on the extent to which the layers in a village are multiplexed.   Specifically, do villages with greater correlation among their network layers experience higher or lower levels of diffusion?
To do this, we first develop a measure of the extent to which a village is multiplexed.

We begin by defining a multiplexing score for household $i$ in village $v$ as
$$
m_{i,v} :=  \frac{\sum_{j}\left(\sum_{\ell}g_{ij,v}^\ell /L\right) }{  \sum_j {\bf 1}\{\sum_\ell g_{ij,v}^\ell>0\}}.
$$
The multiplexing score for a household $i$ measures the average fraction of relationship types it has with each of its neighbors. The numerator is a sum over all neighbors $j$ of the fraction of layers on which household $i$ is connected to $j$. Specifically, for each neighbor $j$, it calculates the fraction of layers connecting $i$ to $j$ by summing the number of links between them across all layers and dividing by the total number of layers $L$, then sums these fractions across all neighbors. The denominator counts the number of unique neighbors of household $i$ by summing an indicator for whether there is at least one link between $i$ and $j$ across any layer. The overall score is thus the average, across neighbors, of the fraction of layers shared with each neighbor.
For example, $m_{i,v}=1$ if whenever household $i$  has a relationship with some other household $j$, then it has all possible relationships with that other household.  
In contrast, when there is no multiplexing, this measure would be $1/L$.

We aggregate this to the village level by taking 
$m_v := \frac{1}{n_v}\sum_i m_{i,v}$. Further, we define a dummy variable for having an above-median amount of multiplexing in the sample as
\[
\text{High Mpx}_v := {\bf 1}\left\{ m_v > \text{median}(m_{1:v})\right\}.
\]

Our regression of interest is 
\begin{align}\label{eq:ols-2}
y_v = \alpha + \beta \cdot DC_v^{advice} \times \text{High Mpx}_v  + \zeta \cdot DC_v^{advice} 
+ \eta \cdot \text{High Mpx}_v +  X_v \Gamma + \epsilon_{v}.
\end{align}
where $DC_v^{advice}$ denotes the diffusion centrality of the seed set in village $v$ for the ``advice" layer (which was singled out as the best predictor of diffusion).

Here, $\zeta$ captures the returns to increasing the diffusion centrality of the seed set. Since information is seeded in all networks, $\eta$ captures how the extent of diffusion changes with the worst possible seeding (the theoretical intercept). The coefficient $\beta$ captures how incrementally improving seeding differentially affects the extent of diffusion as a function of multiplexing. 

The interaction term $DC_v^{advice} \times \text{High Mpx}_v$ is particularly important, and its coefficient of primary interest, since villages with low seed set centrality experience very little diffusion, and hence multiplexing has a very limited opportunity to make any difference in diffusion.   Thus, multiplexing's marginal impact (positive or negative) should be most pronounced in settings where the seed set centrality is high.  

\begin{table}[H]
\caption{Multiplexing and Diffusion}
\label{tab:multidiff}
\centering
\scalebox{0.65}{\begin{threeparttable}
\begin{tabular}[t]{lc}
\toprule
\multicolumn{1}{c}{ } & \multicolumn{1}{c}{Calls per Household} \\
\cmidrule(l{3pt}r{3pt}){2-2}
  & (1)\\
\midrule
High Multiplexing & \num{-0.023}\\
 & \vphantom{1} (\num{0.016})\\
 & {}[\num{0.164}]\\
Seed Set Centrality & \num{0.052}\\
 & (\num{0.016})\\
 & {}[\num{0.002}]\\
High Multiplexing X Seed Set Centrality & \num{-0.039}\\
 & (\num{0.017})\\
 & {}[\num{0.022}]\\
\midrule
Num.Obs. & \num{68}\\
\bottomrule
\end{tabular}
\begin{tablenotes} \footnotesize
\item Robust standard errors are given in parentheses, while p-values are given in square brackets. Seed Set Centrality comes from the "advice" layer and has been standardized. Controls for number of seeds and average total degree across network layers have been added.
\end{tablenotes}
\end{threeparttable}
}
\end{table}

Table \ref{tab:multidiff} reports the coefficient estimates. As expected, the coefficient on seed set centrality is positive and significant.  
We also find that both $\beta < 0$ and $\eta < 0$.  Qualitatively, $\eta<0$ indicates that more multiplexed networks generate less diffusion, with the caveat that these villages could be different for other reasons, and the coefficient is not significant. 
Importantly, the coefficient $\beta < 0$ indicates that villages
with more central seeding---and thus higher levels of diffusion---have their diffusion impeded by multiplexing.


\section{A Theory of Diffusion and Multiplexing}\label{sec:theory}

We now develop a theory that helps us understand how and why multiplexing affects diffusion.  The stylized facts that motivate and structure this theory, established above, are: (i) the network layers are distinct but significantly correlated/multiplexed; (ii) they are differently predictive of diffusion; (iii) multiple layers are predictive of diffusion; 
and (iv) more multiplexed villages experience less information diffusion.

We approach the problem at two levels.
At the individual level, we examine how a node's probability of becoming infected depends on its multiplexing (for any given probability of infection among neighbors).
At the population level, we aggregate the individual effects to analyze broader contagion outcomes.
For this population-level analysis, we use the results about individuals as a key lemma in analyzing a canonical SIS contagion process.

We model two rather different types of processes within a common  framework. The first is ``simple'' diffusion/contagion, in which a single contact is sufficient for an individual to become infected. The second is ``complex'' diffusion, defined as a process in which multiple contacts are needed.  We analyze each type in turn, beginning in each case with a result about individual infection probabilities and then aggregating to the societal level.

We begin by outlining our general model of multiplexed diffusion.


\subsection{A Model of Diffusion with Multiplexing}\label{sec:model}

\

We study diffusion/contagion in a society consisting of a finite set of individuals \(N = \{1, \ldots, n\}\). Each individual has relationships captured via layers \(\{1, \ldots, L\}\), with a generic layer represented by \(\ell\). 
In each layer \(\ell\), the interactions between individuals are described by a (possibly directed) network with adjacency matrix \(g^\ell \in \{0, 1\}^{n \times n}\), such that $g^\ell_{ij} = 1$ if $j$ can transmit to $i$, and $0$ otherwise. (We still speak of the associated edge in the directed graph as being directed from $i$ to $j$, which can be thought of, e.g., as $i$ paying attention to $j$.) We denote the multigraph consisting of $L$ layers by $g = ( g^1, g^2, \ldots , g^L)$. 

Let $\mathcal{L}_{ij} = \{\ell \mid g^\ell_{ij} = 1\}$ denote the set of layers in which there is a directed link from \(i\) to \(j\).
The set of all neighbors for a given node $i$ is denoted  $\mathcal{N}_i = \{j \mid \mathcal{L}_{ij} \neq \emptyset\}$.

To track infection across time, we index discrete periods by $t\in \{0,1,2,\ldots\}$.
At each point in time, an individual in the network is in one of two states: Susceptible (S) or Infected (I). The status of individual \(i\) at time \(t\) is denoted by the random variable \(x_i(t)\). If \(x_i(t) = 1\), individual \(i\) is infected at time \(t\); if \(x_i(t) = 0\), individual \(i\) is susceptible at time \(t\). 
The state of the society at time $t$ is given by the vector \(x(t) = ( x_1(t), x_2(t), \ldots , x_n(t) ) \in \{0, 1\}^n\).

At each time $t$, an individual's state can change based on the infection status of its neighbors. A susceptible individual $i$ becomes infected if it receives at least $\tau$ infection transmissions from its infected neighbors in a given time period.
An infected individual recovers (and becomes susceptible again) randomly with a probability $\delta$ at the end of a period. 
If $\tau=1$, this represents a standard (simple) contagion process, while with a threshold $\tau>1$ this is known as a complex contagion \citep{granovetter1978,centola2010}.\footnote{This is closely related to games on networks \citep{morris2000,jacksonz2014}.}
The greater threshold captures that some behaviors may require more thought, reinforcement, or interactions to become attractive.  For background evidence for such thresholds, there are many such applications as detailed in \cite{centola2010,centola2011,centola2018,boucher2024toward}, for example. 

To complete the description of the model, we examine the mechanics of contagion in more detail. Given that individuals can be connected via multiple layers, we need to define how transmission occurs through multiple layers.
Let \(x_{ij}(t)\) represent the (random) number of infection transmissions at time $t$ to a susceptible node \(i\) from an infected node \(j\), conditional on \(j\) being infected.\footnote{This is related to the modeling of dosed exposures in the literature on contagion; see \citet{dodds2004universal}.} At most one transmission can take place per layer. We denote the distribution of infection transmissions from node $j$ given \(\mathcal{L}_{ij}\) by
\[ f(k; \mathcal{L}_{ij}):= P(x_{ij} = k \; | \; \mathcal{L}_{ij}). \] This is the probability of $k$ transmissions; note
$f(k; \mathcal{L}_{ij})$ can capture arbitrary patterns of correlation in infection transmission through multiple layers.  For each layer \(\ell\), let \(q_\ell\in (0,1)\) be the marginal probability of infection transmission from an infected individual to a susceptible one if they are connected via that layer. 

We emphasize that we allow different layers to have different contact probabilities, which is needed given the heterogeneity in the roles of different layers discussed in Section \ref{sec:determinant_diffusion}.  One can hear information about a new loan program when stopping by to borrow rice from a friend, but it could be more likely that the friend will mention it if one is stopping by to ask for advice.  It could also be that there are two conversations about the loan program, and these could be correlated.  
 If there is a positive correlation in transmission across layers, two nodes connected by layers $\{A,B\}$  have an infection distribution satisfying $f(2,\{A,B\}) \geq q_A q_B$, with negative correlation being the opposite.  

The probability that a susceptible individual \(i\) becomes infected at time \(t\) given the infection status of its neighbors at time \(t-1\) is 
\[ 
P\left(\sum_{j \in \mathcal{N}_i}  x_{ij}  x_j(t-1)\geq \tau \right) \]

\subsubsection{Comparisons of Multiplexing}\label{sec:mpex_comparisons}

\begin{figure}
    \centering
    \begin{subfigure}[b]{0.3\textwidth}
        \centering
        \begin{tikzpicture}[auto,node distance=2cm]
            \begin{scope}
                \node[circle, draw] (1) {1};
                \node[circle, draw, right of=1] (2) {2};
                \node[circle, draw, below of=1] (3) {3};
                \node[circle, draw, right of=3] (4) {4};
                
                \draw[->, red, bend left] (1) to node {} (2);
                \draw[->, green, bend right] (1) to node {} (2);
                \draw[->, blue] (1) to node {} (2);
                
                \draw[->, red, bend left] (1) to node {} (3);
                \draw[->, green, bend right] (1) to node {} (3);
            \end{scope}
        \end{tikzpicture}
        \caption{}
        \label{fig:sub1}
    \end{subfigure}%
    \hspace{0.1\textwidth}%
    \begin{subfigure}[b]{0.3\textwidth}
        \centering
        \begin{tikzpicture}[auto,node distance=2cm]
            \begin{scope}
                \node[circle, draw] (1) {1};
                \node[circle, draw, right of=1] (2) {2};
                \node[circle, draw, below of=1] (3) {3};
                \node[circle, draw, right of=3] (4) {4};
                
                \draw[->, red, bend left] (1) to node {} (2);
                \draw[->, green, bend right] (1) to node {} (2);
                \draw[->, blue] (1) to node {} (2);
                
                \draw[->, red, bend right] (1) to node {} (4);
                \draw[->, green, bend right] (1) to node {} (3);
            \end{scope}
        \end{tikzpicture}
        \caption{}
        \label{fig:sub2}
    \end{subfigure}%
    \hspace{0.1\textwidth}%
    \begin{subfigure}[b]{0.3\textwidth}
        \centering
        \begin{tikzpicture}[auto,node distance=2cm]
            \begin{scope}
                \node[circle, draw] (1) {1};
                \node[circle, draw, right of=1] (2) {2};
                \node[circle, draw, below of=1] (3) {3};
                \node[circle, draw, right of=3] (4) {4};
                \node[circle, draw, left of=1] (5) {5}; 
                
                \draw[->, red, bend left] (1) to node {} (2);
                \draw[->, green, bend right] (1) to node {} (2);
                \draw[->, green, bend right] (1) to node {} (3);
                \draw[->, red, bend right] (1) to node {} (4);
                \draw[->, blue, bend left] (1) to node {} (5);
            \end{scope}
        \end{tikzpicture}
        \caption{}
        \label{fig:sub3}
    \end{subfigure}%
    \caption{{Node 1's relationships are successively less multiplexed moving from panel (A) to (C)}}
    \label{fig:mpex_change}
\end{figure}
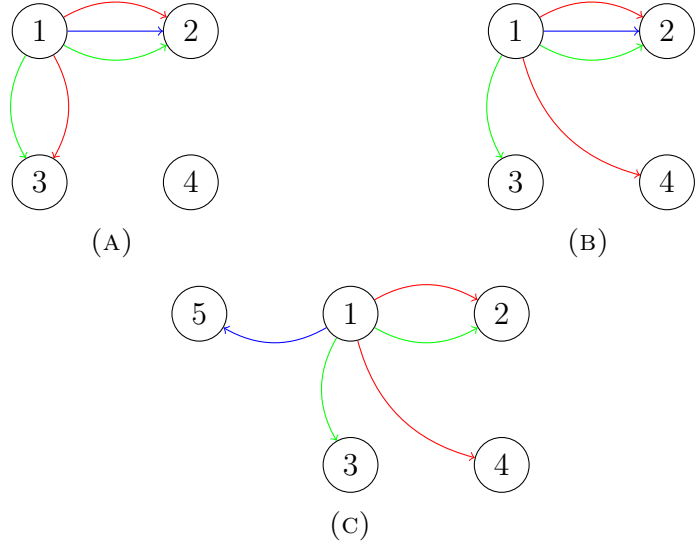

Since it is not always possible to order two multigraphs in terms of multiplexing, we define a partial order on the set of multigraphs. 
We begin with an example illustrating the concept in Figure ~\ref{fig:mpex_change}.

\

In Figure~\ref{fig:mpex_change}(A) we depict a multigraph with 5 nodes and 3 layers. In Figure~\ref{fig:mpex_change}(B), by moving node 1's link in layer \emph{red} from node 3 to node 4, we arrive at a graph that is less multiplexed while maintaining the same out-degree. Similarly, in panel C, we again move node 1's link in layer \emph{blue} from node 2 to node 5, creating a less multiplexed network as compared to panel B.

To formalize this type of ranking, we define a \emph{local multiplexity dominance relation}, denoted by $\prec$. For two multigraphs $g$ and $\widehat{g}$, we say $\widehat{g} \prec g$---that is $\widehat{g}$ is locally less multiplexed than $g$---if $\widehat{g}$ can be obtained from $g$ by removing a link in some layer $\ell$ between nodes $i$ and $j$ and adding a new link in that same layer to another neighbor $k$, where $i$'s connections to $k$ occurred in a set of layers that form a strict \emph{subset} of the layers (except layer $\ell$) in which $i$ was connected with $j$ to start with.   This means that: (i) $\mathcal{L}_{ik}(g)\subsetneq \mathcal{L}_{ij}(g)\setminus \{\ell\} $, (ii) ${g}_{ik}^\ell=0=
\widehat{g}_{ij}^\ell$ and $\widehat{g}_{ik}^\ell=1=
{g}_{ij}^\ell$, and (iii) for all other links $\widehat{g}$ and $g$ coincide.

Given that the \emph{local multiplexity dominance relation} is acyclic (see  Proposition~\ref{prop:graph_acyclic} in the appendix), we define the \emph{less multiplexed relation}, denoted by $\overline{\prec}$ as the transitive closure of $\prec$.  That is, we say that $\widehat{g} \overline{\prec} g$ if there exists a finite sequence of multigraphs $g_1, g_2, \ldots, g_k$ such that $\widehat{g} = g_1 \prec \cdots \prec g_k = g$.
The relation $\overline{\prec}$ forms a partial order on the set of multigraphs.

We now define a corresponding notion for a particular node \( i \). We say that 
\(\widehat{g} \overline{\prec}_i g\) if \(\widehat{g}\) is less multiplexed than \(g\) 
(i.e., \(\widehat{g} \overline{\prec} g\)) and, moreover, the changes in the network’s 
multiplexity structure involve node \( i \). Formally, 
\(\widehat{g} \overline{\prec}_i g\) holds if \(\widehat{g} \overline{\prec} g\) and 
\(\widehat{g}_i \neq g_i\), where \(g_i\) denotes the collection of all layers' adjacency 
for node \( i \). We refer to this refined notion as 
\textit{local multiplexity dominance for node \( i \)}.

\subsection{Multiplexing Impedes Simple Diffusion and Contagion}

\

We first analyze the case of simple contagion, $\tau=1$.
We focus on the case of two layers as this captures all of the essential intuition.

\subsubsection{Infection of an Individual} \,

To understand how increasing multiplexing impedes diffusion, it helps to first isolate 
the comparison on a single pair of links while holding everything else fixed.  Specifically, consider some 
node \( i \) that is connected in both layers \( A,B \) to node \( j \) in \( g \), but in neither layer 
to another node \( k \). Changing from \( g \) to \(\widehat{g}\) involves removing one of the 
layers of \(i\)'s connection to \( j \) and adding it to \( k \). Since all other connections of \( i \) 
remain unaffected, only events involving the changed links need to be considered to assess the effect on $i$'s infection probability.

Suppose that both \( j \) and \( k \) are independently infected with probability \(\rho\), 
and similarly for any of \( i \)'s other connections. The probability of \( i \) becoming infected by one of these two nodes
is higher from two un-multiplexed links if and only if
\[
q_A \rho + q_B \rho - f(2, \{A,B\}) \rho \;\;\leq\;\; 
q_A \rho + q_B \rho - q_A q_B \rho^2,
\]
where \( A,B \) are the layers.\footnote{On the left side, \(q_A \rho + q_B \rho\) represents infection from the single neighbor \(j\) via either layer, and we subtract the overlap where both layers transmit. On the right, the same sum represents infection from two distinct neighbors (\(j\) via layer \(A\) or \(k\) via layer \(B\)), and we subtract the probability that both transmit independently.} Simplifying this yields
\begin{equation}
\label{corr}
q_A q_B \rho \;\leq\; f(2, \{A,B\}).
\end{equation}
A sufficient condition for the inequality is that \(q_A q_B \leq f(2, \{A,B\})\), 
or that transmissions are independent across layers. The basic intuition is that multiplexing 
reduces diversification of contacts across different individuals, which lowers the probability 
of encountering at least one infected neighbor. Under independence or weak correlation, breaking 
a multiplexed link into separate links to distinct neighbors generally improves diffusion.

If there is negative correlation across layers, this condition can be relaxed. As long as $\rho<1$ (so that not everyone is infected), the diversification advantage is preserved even with some negative correlation in transmissions, provided that the  negative correlation 
is not too severe.\footnote{When negative correlation is very strong,  
multiplexing actually enhances simple diffusion processes: having connections in multiple layers to the same neighbor 
disperses the probability of transmission rather than concentrating it. In other words, strong 
negative correlation in transmission events across multiplexed links makes it less likely that one would receive two transmissions from the same neighbor, 
which effectively mimics the benefit of diversified contacts in the independent regime.}

We summarize our observations in the following result.\begin{proposition}\label{prop:individualsimple}
Consider simple contagion ($\tau=1$). If $\widehat g \; \overline \prec_i \; g$ and each of $i$'s neighbors is infected independently with probability $\rho>0$, and $i$ is susceptible, then $i$ is more likely to be infected under the less multiplexed network $\widehat g$ than under $ g$ \emph{if  and only if}
transmission is not too negatively correlated across layers (condition \ref{corr}), with the reverse holding if condition \ref{corr} fails.  
\end{proposition}

\subsubsection{Multiplexing and Overall Infection in the SIS Model}

\

Proposition \ref{prop:individualsimple} gives a sense in which that the infection rate in a variety of contagion processes should be higher on less multiplexed networks. However, our analysis so far only considers one node. We now extend our reasoning to the population level in 
the case of the SIS model. 

To perform this analysis, we extend the mean-field techniques that are standardly used to solve the SIS model with one layer of links (e.g., see \cite{pastor-satorrasv2000,jackson2008}), to study it under multiplexing.  

A given node $i$'s connections are described by a vector $D_i = (D_{i1}, \ldots,D_{iK}) $,
where $K\leq n-1$ is the total number of neighbors of the node, and  $D_{ik}\subseteq \{1,\ldots, L\}$ is the set of layers that $i$ is connected to its $k$th neighbor on,  where each $D_{ik}\neq \emptyset$.

Focusing again on the case of two layers, a sufficient statistic for $D_i$ for the mean-field analysis is a triple $\hat{D}_i=(\hat{D}_{iA},\hat{D}_{iB},\hat{D}_{i,AB})$, 
which represents the number of connections that $i$ has that are just on layer $A$, just on layer $B$, and on both layers, respectively.
The distribution of $\hat{D}$ across the population is described by a function $P(\hat{D})$ that has finite support.
The steady-state infection rate of nodes with connection profile $\hat{D}$ is denoted $\rho(\hat{D})$. 
The population infection rate is then defined by
\begin{equation}
\label{rho}
\rho=\sum_{\hat{D}} P(\hat{D})\rho(\hat{D}). 
\end{equation}

The probability that a susceptible node with connections 
$\hat{D}=(\hat{D}_A,\hat{D}_B,\hat{D}_{AB})\neq (0,0,0)$  becomes infected, in steady state, is then\footnote{Here, $(1-\rho q_A)$ is the probability that a layer-$A$-only neighbor fails to transmit infection, and similarly for $(1-\rho q_B)$ for a layer-$B$-only neighbor. For neighbors connected via both layers, 
$(1 - \rho [q_A + q_B - f(2,\{A,B\})])$ is the probability that such a neighbor fails to transmit 
infection, accounting for potential correlation in transmissions across the two layers.
Raising these terms to the powers $\hat{D}_A, \hat{D}_B, \hat{D}_{AB}$ accounts for all relevant neighbors. 
Multiplying them together gives the probability that \emph{none} of these neighbors transmit 
infection through their respective sets of layers. Subtracting this product from 1 then yields the 
probability that at least one transmission succeeds, infecting the susceptible node.}
\begin{equation}
\label{sis}
1- (1-\rho q_A)^{\hat{D}_{A}}(1-\rho q_B)^{\hat{D}_{B}} (1- \rho [q_A  + q_B - f(2, \{A,B\})])^{\hat{D}_{AB}}.
\end{equation}

In the mean-field analysis, the steady state equation for nodes with connections
$\hat{D}=(\hat{D}_A,\hat{D}_B,\hat{D}_{AB})\neq (0,0,0)$, as a function of the overall infection rate $\rho$,
is the solution to 
\begin{equation}
\label{sis-steady}
\rho(\hat{D}) \delta = 
\end{equation}
$$
(1- \rho(\hat{D})) \left[  1- (1-\rho q_A)^{\hat{D}_{A}}(1-\rho q_B)^{\hat{D}_{B}} (1- \rho  (q_A  + q_B - f(2, \{A,B\})))^{\hat{D}_{AB}}\right].
$$

A steady-state is a joint solution to (\ref{rho}) and (\ref{sis-steady}) for each $\hat{D}$ in 
the support of $P$.  Note that $0$ is always a solution, and for some distributions $P$ there may also 
exist a positive solution. We focus on the largest positive solution, which is the one that corresponds to the behavior of large finite graphs.\footnote{See \citet{elliott2022supply} for a detailed argument in an analogous situation.}

We extend the partial order we defined in \ref{sec:mpex_comparisons} to the space of distributions as follows. We say that a distribution ${P}'$ is less multiplexed than $P$,  denoted by $ P' \overline \prec  P$, if there exists $\hat{D}$ and
$\hat{D}'$ such that 
\begin{itemize}
    \item $\hat{D}'_A = \hat{D}_A+1 $, 
    \item $\hat{D}'_B = \hat{D}_B+1 $,
    \item $\hat{D}'_{AB} = \hat{D}_{AB}-1 $,
\item $P'(\hat{D}')+ P'(\hat{D}) = P(\hat{D}')+ P(\hat{D}) $,
and
\item $P'(\hat{D}') > P(\hat{D}')$.
\end{itemize}
In other words, to move from \(P\) to \(P'\), we increase the frequency of profiles with separate 
links \((\hat{D}'_A,\hat{D}'_B)\) while reducing the frequency with multiplexed links \((\hat{D}'_{AB})\), 
holding total mass constant. The relation \( \overline{\prec} \) is then defined as the transitive 
closure of this ordering.

\begin{proposition}
\label{prop:sis-simple}
Consider a simple contagion process ($\tau=1$) process. Let transmission probabilities be given by $f$ with marginal probabilities $(q^\ell)_\ell \in (0,1)^L$. Finally, fix a recovery rate $\delta\in (0,1)$ and two distributions of connections 
$ P'$ and $ P$ that each have positive steady-state infection rates.
    If  $ P' \overline \prec  P$, then the positive steady-state infection rate under $P'$ is higher than that under $P$ \emph{if and only if} transmission is not too negatively correlated (condition \ref{corr}) at the positive infection rate of $P$.
\end{proposition}

Proposition \ref{prop:sis-simple} implies that multiplexing has significant consequences, which can be beneficial or detrimental depending on whether diffusion is socially desirable  (e.g., information about a beneficial program) or not (e.g., spread of a disease).
Given the various factors that may lead to multiplexing, this implies that the mechanisms causing people to layer their networks have important implications for diffusion processes.  
This also means that networks whose layers are optimized for one purpose may be suboptimal for another.

\subsection{Multiplexing and Complex Diffusion}\label{sec:complex_cont}

\ 

The results on simple contagion are unambiguous: multiplexing impedes simple diffusion/contagion except in extreme cases of negatively correlated transmission probabilities. 
Complex contagion, in contrast, presents a more nuanced picture.  Multiplexing can both enhance and impede diffusion, depending on the circumstances. 

In complex diffusion, two competing forces of multiplexing emerge. One force mirrors the effect seen in simple contagion: diversifying links increases the probability of at least some links reaching infected individuals. However, a counterforce now exists: conditional on reaching an infected individual, multiplexing leads to higher probabilities of multiple transmissions, compared to spreading those links across other individuals who might be uninfected. This makes it more likely that a contagion threshold greater than $1$ is reached.

To keep the analysis as uncluttered as possible, we again focus on the case of two layers. We also consider a case where the correlation in transmission across layers is nonnegative and not too high; formally,there is an $\varepsilon$, to be determined in the proofs of the propositions below,
for which $(1+\varepsilon) q_A q_B \geq f(2,\{A,B\})\geq q_Aq_B$.  Of course, a sufficient condition
for this to hold is independent transmission, under which $f(2,\{A,B\})= q_A q_B$.   This condition is needed as with excessive positive or negative correlation in transmission, strange discrete behavior in transmission as a function of multiplexing can occur.\footnote{For instance, if transmission is perfectly positively correlated, then one is always more likely to get two transmissions from a single multiplexed connection than two unmultiplexed connections, but is always more likely to get one transmission from the reverse.  This then implies that the optimal configuration of connections depends on whether $\tau$ is even or odd, in complicated ways as a function of a node's overall degrees in each layer.}  The restriction to two layers allows the results to highlight the more fundamental forces of multiplexing.

\begin{proposition}\label{prop:individualcomplex}
Consider a complex contagion ($\tau>1$).  
Fix a susceptible node $i$ such that $i$'s neighbors are infected independently with probability $\rho>0$, and two networks such that $\widehat g \; \overline \prec_i \; g$.  
Also suppose that $\sum_{\ell, j} g_{ij}^\ell>\tau$, so that $i$ has more than enough connections to become infected. 

There exist $0<\underline{\rho}<\overline{\rho}<1$ such that 
\begin{itemize}
\item  if $\rho, q_A, q_B >\overline{\rho}$, then $i$ is \emph{less} likely to be infected under the more multiplexed network $g$ than under $\widehat g$, and
\item  if $\rho, q_A, q_B <\underline{\rho}$, then $i$ is \emph{more} likely to be infected under the more multiplexed network $g$ than under $\widehat g$. 
\end{itemize}    
\end{proposition}

The intuition behind this result is as follows. There exist nodes $i,j,k$ such that under $g$, node $i$ is connected to $j$ on two layers and to $k$ on none, while under $\widehat g$, node $i$ is connected to $j$ on one layer and to $k$ on the other.  
The cases in which this difference can be pivotal are when the other connections to other nodes have led to either $\tau-1$ or $\tau-2$ transmissions.
With high infection and transmission rates among neighbors, the $\tau-1$ case predominates, making the situation resemble simple contagion---thus, less multiplexing leads to a higher chance of infection. Under low infection rates, the $\tau-2$ case becomes more likely, requiring two incremental infections. This is highly improbable across two separate neighbors but more likely with a single neighbor, making more multiplexing advantageous for infection probability.

Interestingly, as we will see in the simulations below, these forces can interact non-monotonically in the intermediate range for infection and transmission rates, which explains the gap between the upper and lower bounds.

We now state how this translates into an aggregate infection rate.

\begin{proposition}\label{prop:SIScomplex}
Consider a complex contagion ($\tau>1$), with nonnegative correlation in transmission across layers, so that $f(2,\{A,B\})\geq q_A q_B$,  and two distributions such that $P' \overline \prec P$ and both have positive steady-state infection rates.   Also suppose that $\hat{D}_A+\hat{D}_B+2\hat{D}_{AB}>\tau$ for each $\hat{D}$ in the distribution $P$, so that each node has more than enough connections to become infected. 
There exist $0<\underline{\rho}<\overline{\rho}<1$ such that 
\begin{itemize}
    \item  if $q_A,q_B<\underline{\rho}$ and $\delta$ is sufficiently high, then the steady-state infection is higher for $P$ than $P'$, and 
    \item  if $q_A,q_B>\overline{\rho}$ and $\delta$ is sufficiently low, 
    then the steady-state infection is lower for $P$ than $P'$. 
\end{itemize}    
\end{proposition}

Note that the steady-state infection of every connection type shares the same ordering as the overall infection rate.

In terms of the reflection of the theory on the empirical results, the fact that more multiplexed villages had lower diffusion conditional on enough seeds to get the process going   suggests that either the process was simple diffusion (which would be consistent with simply needing to know about the free phone giveaway), or that it had a relatively low threshold in most people and there was enough participation so that spreading links out was more likely to lead to a pivotal contact.  
As we do not fully observe the information spread in the villages, to get a deeper understanding of how the diffusion works in those villages
we next do some simulations on those village networks to illustrate the theoretical results and explore some of the comparative statics.

\subsection{Simulations}

To get a deeper feeling for how the theory depends on the details of the process, 
we run simulations on the networks from the RCT villages. 
We simulate a Susceptible-Infected-Susceptible (SIS) diffusion process for the cases of both simple and complex diffusion and compare outcomes as multiplexing is varied.
In order to compare across similar-sized networks where only multiplexing is changing, we take a given village network and construct two-layer networks by combining different pairs of empirical networks (which end up empirically having different multiplexing rates), in a way we specify below.  We then perform many diffusion simulations on these two-layer networks for each village.

More specifically, for each village, we begin by picking three empirical adjacency matrices representing different network layers sorted in decreasing order of their average out-degree: $A_1$, $A_2$, and $A_3$.  We then pair $A_1$ with $A_2$ for one simulated diffusion, and $A_1$ with $A_3$ for the other.  To ensure that the average out-degree is comparable across the networks, we prune at random the links in $A_2$ to match the average out-degree of $A_3$, resulting in a pruned network $A_2^{'}$. We construct two multiplexed networks: $g'$, by combining $A_1$ and $A_2^{'}$, and $g$, by combining $A_1$ with the $A_3$. The process is presented in full detail in the Appendix (Algorithm \ref{alg: graph_rewire}).
 
The diffusion process (also presented in full detail in the appendix in Algorithm \ref{alg: graph_diffusion}), is as follows. First, a susceptible node can get message transmissions from each infected neighbor in each layer, i.i.d., with probability $q$ in each period. Second, a susceptible node gets infected only if it receives at least $\tau\geq 1$ contacts in a given time period and the count resets in each time period. Third, in each period an infected node transitions back to being susceptible with probability $\delta$. We terminate the simulation when the share of infected nodes changes by less than a small threshold between consecutive iterations. 
In our simulations, we use $\tau =1$ for simple diffusion and $\tau=2$ for complex.  
In each simulation we set the number of randomly selected seeds in the initial period to be $\floor{\sqrt{n}}$, where $n$ is the number of households in the network. 
For both, simple diffusion ($\tau=1$) as well as complex diffusion ($\tau=2$) we run simulations on a grid of $(q,\delta) \in [0.1,0.5]\times [0.1,0.5]$.
We run the diffusion simulations $500$ times
for each village across both multiplexed networks $g,g'$ described above. We report the averages across all $70$ villages.

Given that these are smaller networks, some simulations end up randomly having more or less diffusion in any given run across the two comparison networks.  Thus, we
tabulate the fraction of simulation runs for which more multiplexing is associated with more diffusion.

\begin{figure}[h]
    \centering
    \begin{subfigure}[t]{0.48\linewidth}
        \includegraphics[width=\linewidth]{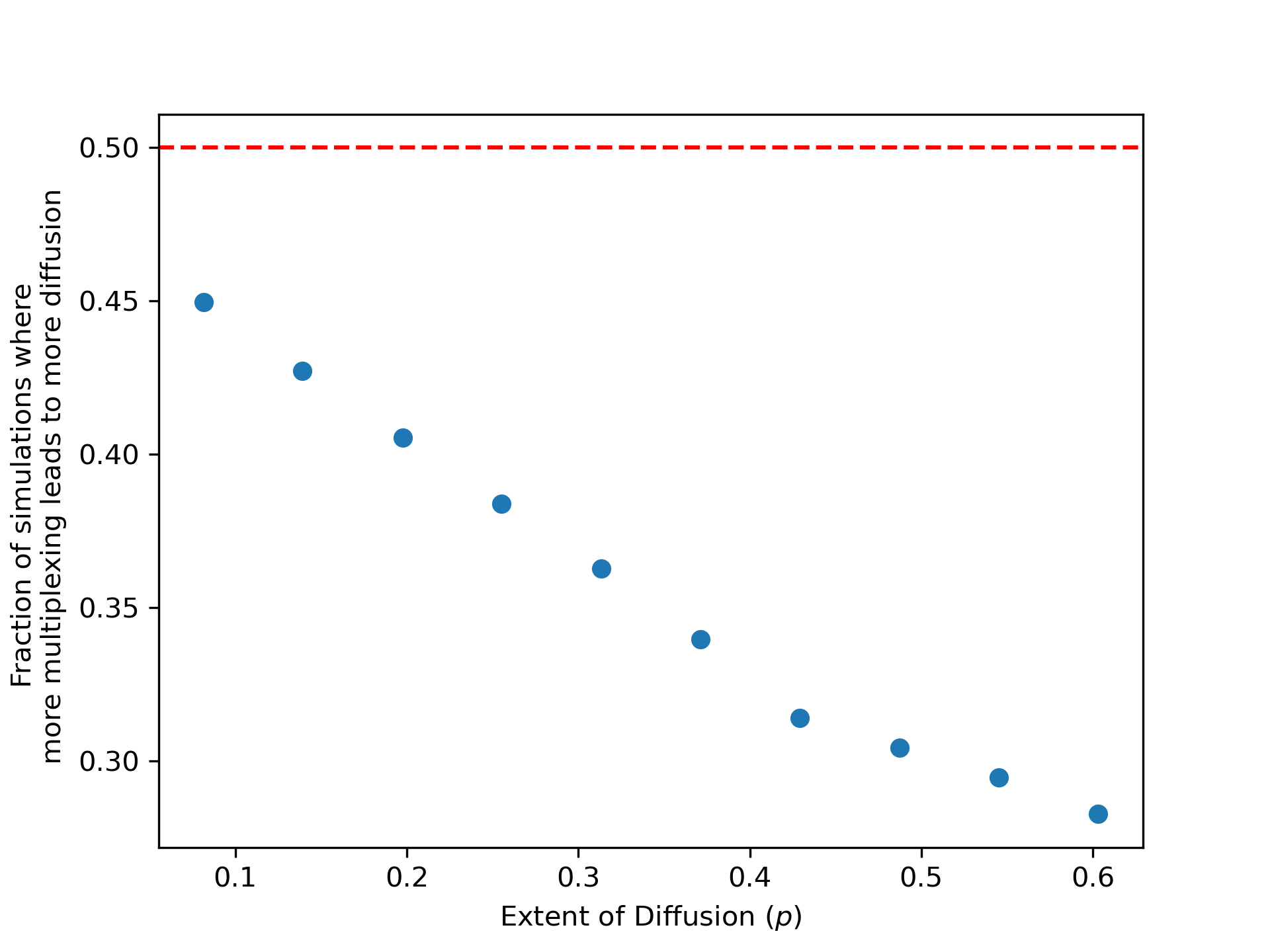}
        \caption{Simple Contagion ($\tau = 1$)}
    \end{subfigure}
    \hfill
    \begin{subfigure}[t]{0.48\linewidth}
        \includegraphics[width=\linewidth]{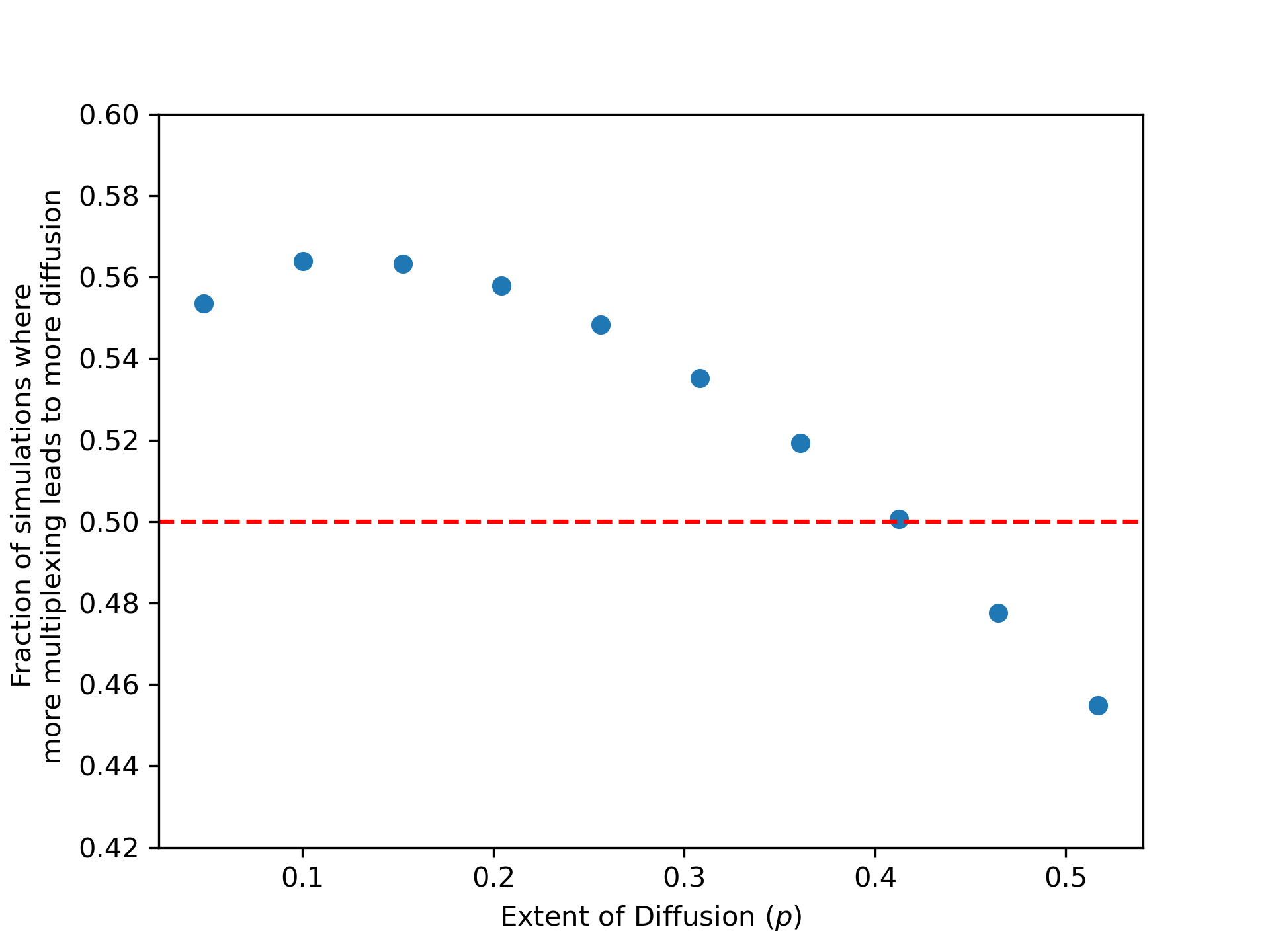}
        \caption{Complex Contagion ($\tau = 2$)}
    \end{subfigure}    
    \caption{Diffusion Simulations}
 	\label{fig:contagion}
\end{figure}

In Figure \ref{fig:contagion} we plot the fraction of simulation runs where more multiplexing leads to more diffusion against the extent of diffusion in the network $p$. 
In panel A, we plot the results for simple diffusion. We find that higher multiplexing is consistently associated with lower diffusion levels, as in our theoretical results.

In panel B we see the nonmonotonicity from the countervailing forces in complex diffusion that we mentioned in Section \ref{sec:complex_cont}.
We also see a confirmation of the theoretical results.
At low levels of diffusion, the steady state diffusion is increasing in multiplexing, and for high diffusion levels, the steady state diffusion is decreasing in multiplexing.


\section{Discussion and Further Observations}
\label{sec:discussion}

Our study began by examining patterns of multiplexing in two large data sets.  We next showed that multiplexing systematically impacts diffusion, via both experimental evidence and theoretical modeling.

Our findings highlight the need for future work on incentives to multiplex and the consequences of multiplexing decisions. There are several immediate directions to explore. For example, our results suggest that the deeper the need to form reinforced or supported (i.e., multiplexed) relationships, the greater the potential inefficiencies in certain domains. In particular, those who are under weaker institutions or have limited resources may face a greater need to multiplex relative to their richer counterparts. Consequently, they may experience both reduced access to information and increased susceptibility to the spread of social norms that are described by complex contagion dynamics---a susceptibility that may be beneficial or detrimental.

There is also a need for further development of measures and methods of analyzing multiplexed networks.  We defined one of many potential measures of how multiplexed a network is, as well as one of many potential partial orders. Understanding which measures are most appropriate in which settings is a subject for further research.

To close, we report two other patterns that we found in the data.
For both of the following calculations, 
we use the multiplexing score that we defined in Section \ref{defmultiplex}:
$$
m_{i,v} :=  \frac{\sum_{j}\left(\sum_{\ell}g_{ij,v}^\ell /L\right) }{  \sum_j {\bf 1}\{\sum_\ell g_{ij,v}^\ell>0\}},
$$
where $i$ represents either an individual or a household, depending on the analysis.

\begin{figure}[t]
    \centering
    \includegraphics[scale = .8]{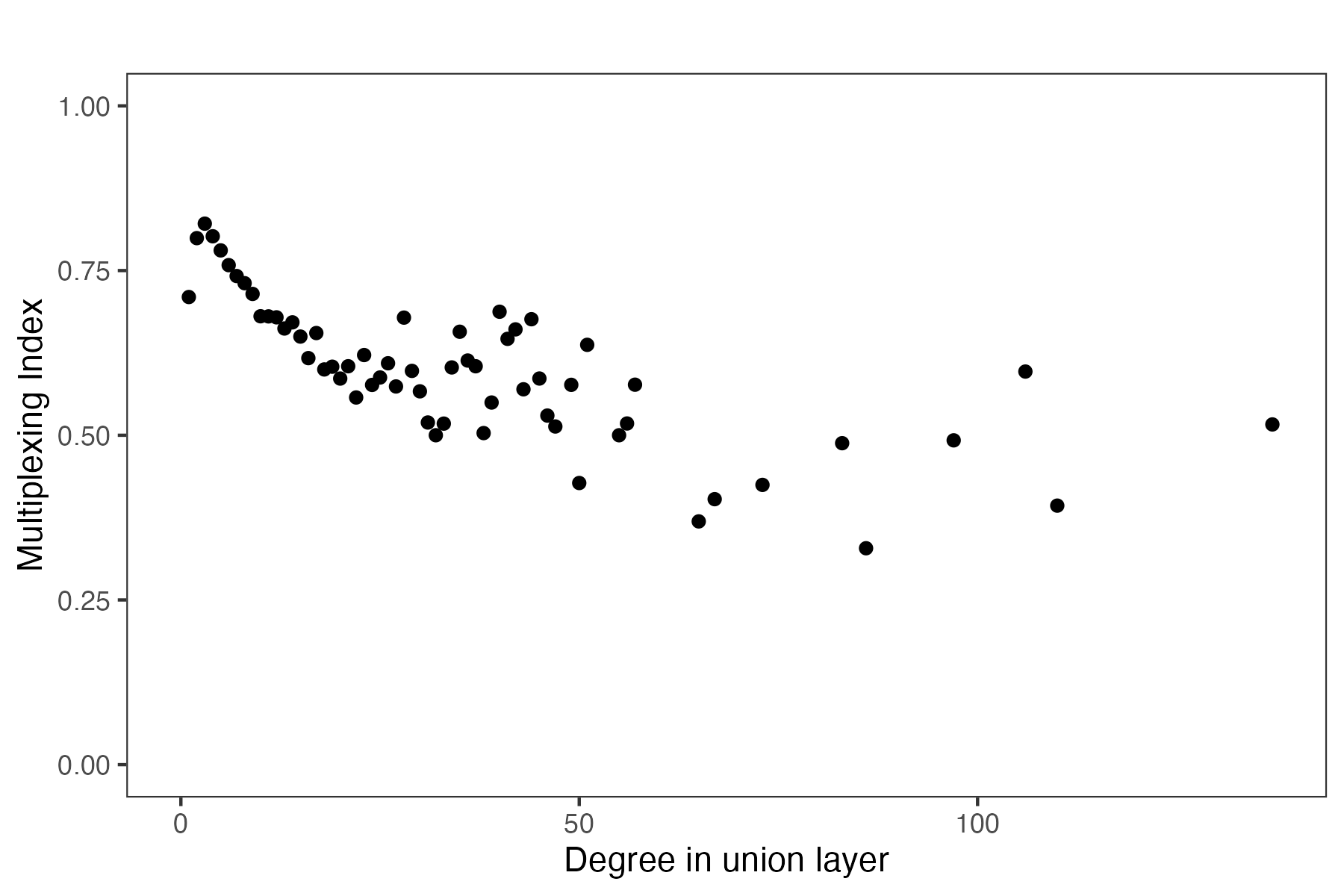}
    \caption{Multiplexing as a function of degree.}
    \label{fig:deg_vs_mpex}
\end{figure}

The first pattern is that higher-degree households are less multiplexed. We restrict our attention to the elicited layers in the RCT villages: the social, kerorice, advice, and decision layers. Figure \ref{fig:deg_vs_mpex} depicts a binned scatter plot where we can see that households that have higher degree (aggregated across layers) have lower levels of multiplexing.

\begin{figure}[H]
	\centering
	\subfloat[Density: Aggregate]{%
		\includegraphics[width=0.47\linewidth]{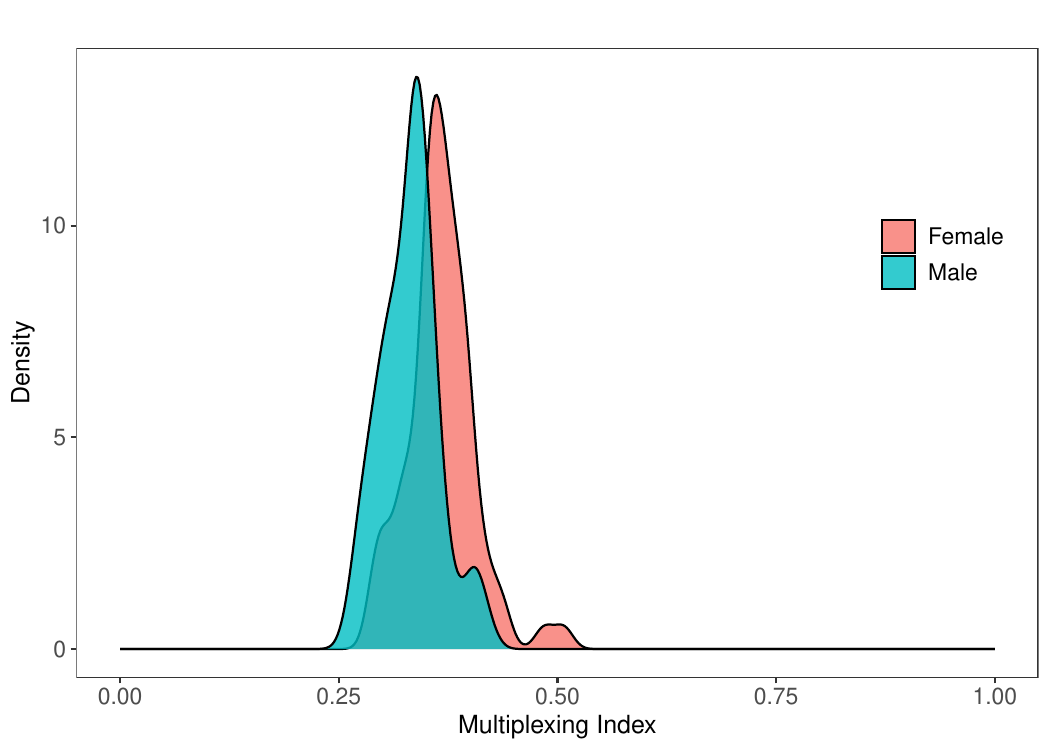}
	}
	\hfill
	\subfloat[Cumulative Density: Aggregate]{%
		\includegraphics[width=0.47\linewidth]{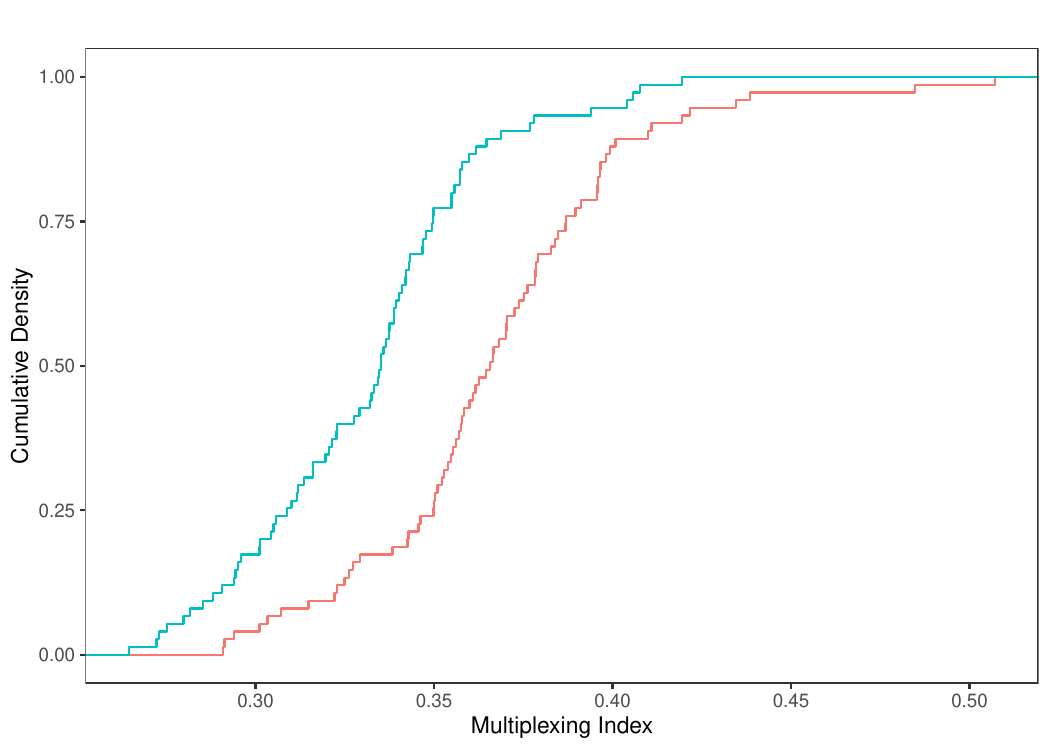}
	}
	\\
	\subfloat[Density: Individual]{%
		\includegraphics[width=0.47\linewidth]{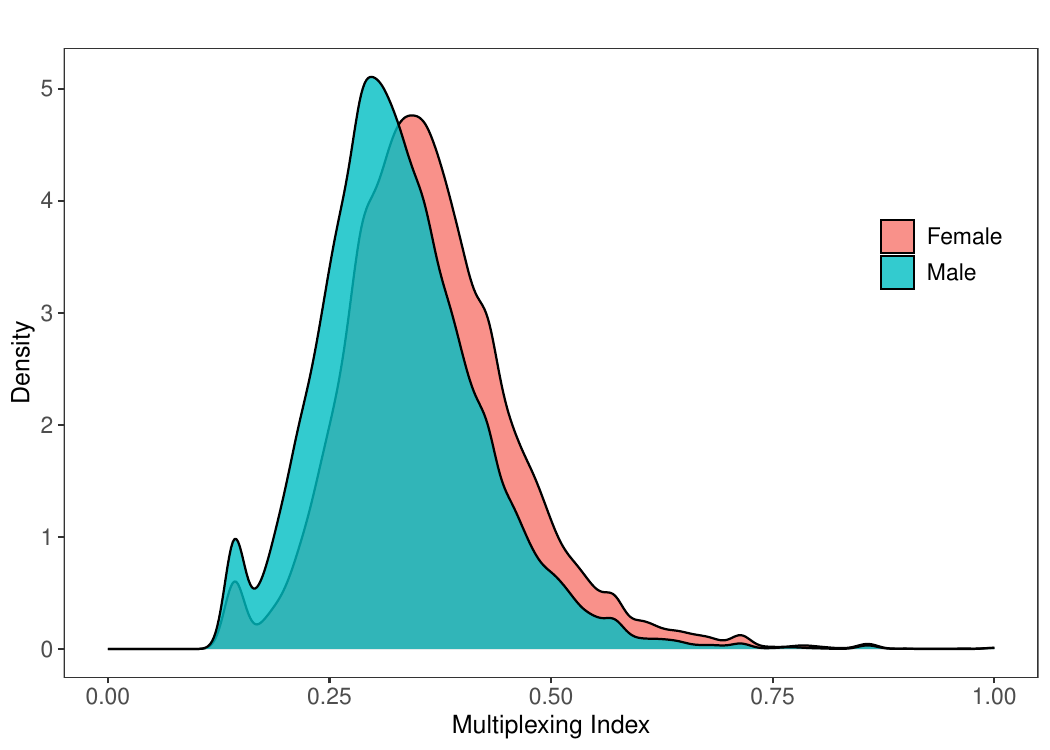}
	}
	\hfill
	\subfloat[Cumulative Density: Individual]{%
		\includegraphics[width=0.47\linewidth]{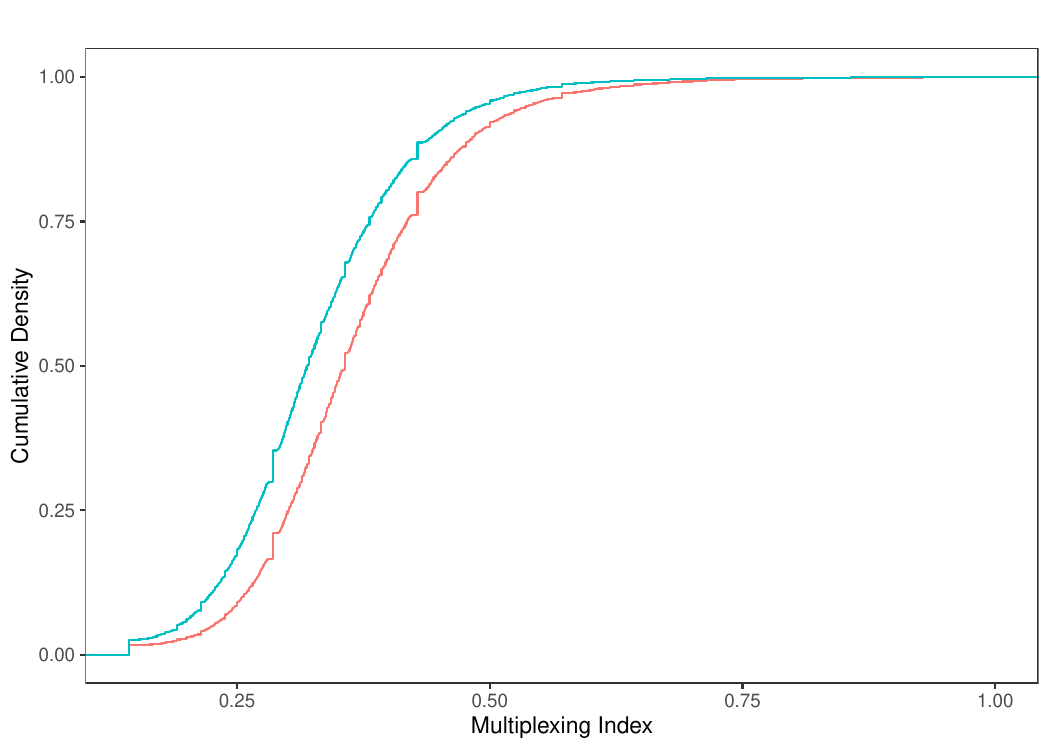}
	}
	\caption{Multiplexing by Gender. The aggregate plots average over a given gender within a village and then depict the distribution of the resulting numbers for that gender. The individual plots include each person as a separate observation.}
	\label{fig:mpex_gender}
\end{figure}

The second pattern is that women's networks are significantly more multiplexed than those of men. Here we use the microfinance villages, where we have access to individual-level network data. 
We focus on the social, kerorice, advice, decision, money, temple, and medic layers. For each village $v$, we aggregate this score at the gender level: $m_{a,v} = \frac{1}{n_a} \sum_{i \in I_a} m_{i,v}$, where $a \in \{\text{male},\text{female}\}$. 
Figure \ref{fig:mpex_gender} shows the density curves for these multiplexing scores across the villages, as well as for each individual treated as a separate observation.  The distributions reveal that women's networks are systematically more multiplexed.
In Supplemental Appendix Figure \ref{fig:mpex_gender_app} we include the same analysis with a different wave of data, and see an even starker difference.

This result could help explain results of \cite{beaman2018diffusion}, who found unexplained differences in diffusion by gender. To understand potential sources of gender differences in multiplexing, note that women in rural Indian communities often marry across village boundaries (though frequently still within the constraints of caste/jati endogamy) and most of these marriages are virilocal---requiring the wife to move into the husband's house \citep{rosenzweig1989consumption, rao2015marriage}. As a consequence, women often rely on affinal kin and over time need to ``rebuild'' their networks \citep{hruschka2023starting}. This occurs in conjunction with the expectation that these women take on various responsibilities, including agricultural work, managing the household, preparing meals, and raising children. Such constraints on available relationships while serving multiple roles can plausibly result in
high levels of multiplexing, an interesting subject for further research.

\bibliographystyle{ecta}
\bibliography{multiplex}


\newpage

\setcounter{table}{0}
\renewcommand{\thetable}{S\arabic{table}}

\setcounter{figure}{0}
\renewcommand{\thefigure}{S\arabic{figure}}

\appendix

\centerline{\bf Supplemental Online Appendix for}
\centerline{\bf Multiplexing in Networks and Diffusion}
\centerline{\bf by Chandrasekhar, Chaudhary, Golub, Jackson}

\section{Additional Data Background}\label{appdata}

\subsection{Descriptive Statistics}

Descriptive statistics are 
presented in Table \ref{table:desc}.  The different layers exhibit significantly different patterns of connection.   For example, in both datasets, the social layer is denser than the other layers and has among the highest levels of clustering.  We observe a higher variance of node degrees in the decision layer than in other comparable layers (e.g., advice).%
\footnote{In the RCT villages the social layer is significantly denser than kerorice, advice, or decision layers (p-values $0.009$, $0.000$, and $0.000$ respectively). The advice layer has significantly less variance (p-value $0.0006$) and more clustering (p-value $0.03$) than the decision layer. Similar patterns hold in the Microfinance villages.}  

\begin{table}[h!]
\caption{Descriptive Statistics\label{table:desc}}

\begin{tabular}[t]{lrrrrr}
\toprule
Network & degree & degree S.D. & density & triangles & clustering\\
\midrule
\addlinespace[0.3em]
\hline
\multicolumn{6}{l}{\textbf{Microfinance villages}}\\
\hline
\hspace{1em}social & 15.296 & 7.841 & 0.079 & 2635.040 & 0.252\\
\hspace{1em}kerorice & 7.029 & 3.834 & 0.037 & 594.160 & 0.259\\
\hspace{1em}advice & 6.158 & 3.835 & 0.032 & 299.120 & 0.168\\
\hspace{1em}decision & 6.553 & 4.309 & 0.034 & 356.040 & 0.169\\
\hspace{1em}money & 8.512 & 5.036 & 0.044 & 681.960 & 0.193\\
\hspace{1em}temple & 1.709 & 1.899 & 0.009 & 52.040 & 0.175\\
\hspace{1em}medic & 6.530 & 3.911 & 0.034 & 369.400 & 0.188\\
\hspace{1em}union link & 75.428 & 32.542 & 0.368 & 314121.027 & 0.862\\
\hspace{1em}intersect link & 0.576 & 0.883 & 0.003 & 7.000 & 0.203\\
\hspace{1em}jati & 68.291 & 34.293 & 0.332 & 310150.907 & 1.000\\
\addlinespace[0.3em]
\hline
\multicolumn{6}{l}{\textbf{RCT villages}}\\
\hline
\hspace{1em}social & 5.711 & 3.626 & 0.031 & 251.271 & 0.185\\
\hspace{1em}kerorice & 4.910 & 3.235 & 0.027 & 176.557 & 0.174\\
\hspace{1em}advice & 4.197 & 3.091 & 0.023 & 124.100 & 0.161\\
\hspace{1em}decision & 4.206 & 3.675 & 0.023 & 125.571 & 0.145\\
\hspace{1em}union link & 55.756 & 27.861 & 0.296 & 150771.400 & 0.913\\
\hspace{1em}intersect link & 1.812 & 1.829 & 0.010 & 38.871 & 0.229\\
\hspace{1em}jati & 52.633 & 28.599 & 0.279 & 150117.500 & 1.000\\
\bottomrule
\end{tabular}

\end{table}

We also observe that the microfinance villages and RCT villages differ from each other in some descriptive statistics. Microfinance villages across all network layers are denser on average and exhibit higher levels of clustering, as can be seen in Table \ref{table:desc}. The two samples also slightly differ in terms of village size. RCT villages have 197 households on average, while microfinance villages are larger with 216 households on average.

Additionally, the jati layer has by far the highest degree. This finding foreshadows that jati match serves as a poor proxy for other types of relationships, being too dense, too clustered, and too homophilous to predict the other layers.

\subsection{Additional RCT Details}

In particular, they had to dial the provided promotional number and leave a ``missed call.''  This was a call that we registered but did not answer and was free for the participant to make, which was a standard technique for registration at the time.
Registered callers were visited a few weeks later and received a reward.
The individual rolled a pair of dice and received INR 25 $\times$ the number rolled. This yielded cash prizes of amounts ranging from INR 50 (for a 2) to INR 275 (for an 11). A roll of 12 was rewarded with a cell phone worth INR 3000. The expected value of the prize was INR 255, which was more than half of a day's wage in the area.

\subsubsection{Network Layers}
\

In terms of notation, we define a multi-layered, undirected network for each village $v$, for layer $\ell = 1,\ldots , L$, with $g_{ij,v}^\ell = 1$ if either household $i$ or $j$ reported having a relationship of type $\ell$. We add another layer where $i$ and $j$ are linked if they belong to the same jati.  For the Microfinance Village Sample, where GPS data are available, we construct a weighted graph where the $ij$ entry is the geographic distance between the two households. 

We construct three \emph{synthetic network} layers. The
\emph{union} layer has a link is present if a link exists in any layer. The \emph{intersection} layer has a link is present if it exists in all layers. Finally, the \emph{total} network is constructed as a weighted and directed network whose edge weights are the sums of indicators for links in all directed layers (using the raw directed nomination matrices, thus excluding jati and geography).

\subsection{Principal Component Analysis (PCA)} \label{app:pca}
\

We perform a principal component analysis with all of the layers (excluding the synthetic \emph{union}, \emph{intersection} layers and \emph{total} network layers).
We treat each pair of households (in a given village) as an observation, yielding $\sum_v \binom{n_v}{2}$ observations, where $n_v$ is the number of households in village $v$, and the number of dimensions is the number of layers $L$ in the given sample. 

\subsubsection{The Backbone Network Construction}
\

The backbone network is built using the first $K$ principal components derived from the PCA. To select the optimal $K$ number of principal components the literature usually relied on a cutoff based on patterns of either decreasing eigenvalues or increasing variability of eigenvectors. \cite{luoli2016laddle} combine these two approaches to better estimate the optimal $K$. They propose a new estimator, called the ``ladle estimator'' which minimizes an objective function that incorporates both the magnitude of eigenvalues and the bootstrap variability of eigenvectors. This approach exploits the pattern that when eigenvalues are close together, their corresponding eigenvectors tend to vary greatly, and when eigenvalues are far apart, the eigenvector variability tends to be small. By leveraging both sources of information, the ladle estimator can more precisely determine the rank of the matrix, and thus the optimal number of components to retain.

For a pair $ij$ in village $v$, we compute the weighted sum of its projections on the first $K$ principal components as 
\[
Z_{ij,v} = \sum_{k=1}^K w_k \cdot \left(\sum_{\ell=1}^L g^{\ell}_{ij,v} \cdot e_{k\ell} \right).
\]
In this formula, $e_k$ is the eigenvector associated with the $k^{th}$ principal component, and the weights $w_k$ are determined by the relative magnitudes of the eigenvalues associated with each component:
\[
w_k :=  \lambda_k / \sum_{j=1}^K \lambda_j.
\]

For each village $v$, we then define a ``backbone'' network, $g^{\text{backbone}}$, from the principal components as a weighted graph where 
\[
g_{ij,v}^{\text{backbone}} = Z_{ij,v}.
\]

\subsection{Variable definitions}\label{app:var_def}

\subsubsection{Diffusion Centrality}
\

We use a specific diffusion centrality measure developed in \cite{banerjeecdj2013} and further studied in \cite{banerjeecdj2019}. 
In particular, the \emph{diffusion centrality} of a node
$j$ in layer $\ell$ in village $v$, $DC_{j,v}^\ell$
is defined by
\[
DC_{j,v}^\ell := \left[\sum_t^T (qg^\ell_v)^t \cdot 1 \right]_j,
\]
where $T$ is the number of rounds of communication and $q$ is the probability of transmission in each period across any given link.
Following \cite{banerjeecdj2019}, for village $v$ and network layer $\ell$, we set $T = \operatorname{diameter}(g_v^\ell)$, and $q = 1/\lambda_v^\ell$, where $\lambda_v^\ell$ is the largest eigenvalue associated with $g_v^\ell$.
\footnote{See \cite{banerjeecdj2019} for a theoretical foundation for using these as default settings in diffusion centrality.} 
We calculate the diffusion centrality of the seed set of village $v$, $S_v$, for layer $\ell$ by
\[
DC_v^\ell := \sum_{j\in S_v} DC_{j,v}^\ell.
\]

\subsubsection{The Multiplexing Score}
\ 

We define a multiplexing score for household $i$ in village $v$ as 

$$
m_{i,v} :=  \frac{\sum_{j}\left(\sum_{\ell}g_{ij,v}^\ell /L\right) }{  \sum_j {\bf 1}\{\sum_\ell g_{ij,v}^\ell>0\}}.
$$

This measures the average fraction of relationship types household $i$ has with each of its neighbors. The numerator calculates the average number of links household $i$ has to each neighbor across all $L$ relationship types. It does this by first summing the number of links between household $i$ and each neighbor $j$ across all layers, dividing by the total number of layers $L$, and then summing this average across all neighbors $j$. The denominator counts the number of unique neighbors of household $i$ by summing an indicator for whether there is at least one link between $i$ and $j$ across any layer.

We aggregate this to the village level by taking 
$m_v := \frac{1}{n_v}\sum_i m_{i,v}$. Further, we define a dummy variable for having an above-median amount of multiplexing in the sample as
\[
\text{High Mpx}_v := {\bf 1}\left\{ m_v > \text{median}(m_{1:v})\right\}.
\]

\subsection{LASSO}\label{app:lasso}
\

To select a sparse set of network layers that best predict diffusion we perform a LASSO ($\ell_1$ penalized) regression. The regression of interest is given by 

The regression of interest is given by
$$
y_v = \alpha + \sum_\ell \beta^\ell \cdot DC_v^\ell +  X_v \Gamma + \epsilon_{v}
$$

where $DC_v^\ell$ represents the seed set diffusion centrality for layer $\ell$ in village $v$. We are interested in which $\beta^\ell$ are estimated to be non-zero and the consistent estimates of these parameters.

Given the high correlation between network layers, we fail to satisfy the irrepresentability condition which requires that the regressors of interest not be excessively correlated \citep{zhao2006model}. To overcome this problem, we use the Puffer transformation developed by \cite{rohe2015preconditioning} and  \cite{jia2015preconditioning}, which recovers irrepresentability when the number of observations exceeds the number of variables.
Although the regressors $(DC^\ell_v)_{v,\ell}$ have correlated columns, by appropriately pre-conditioning the data matrix,
we can force its columns to be orthogonal and therefore irrepresentable. Puffer-LASSO then recovers the set of relevant variables with probability tending to one exponentially fast in the number of observations, with consistent parameter estimates, that are asymptotically normally distributed with probability approaching one \citep{javanmard2013model,jia2015preconditioning,taylor2015statistical,lee2016exact,banerjee2021selecting}.



\section{Proofs}\label{proofs}

\

\noindent {\bf Proof of Proposition \ref{prop:individualsimple}}:
We adopt the notation from Proposition \ref{prop:sis-simple}, as given independent probabilities of infection of neighbors,
the probability that an individual
with connection profile $\hat{D}=(\hat{D}_{A},\hat{D}_{B},\hat{D}_{AB})$ on network $g$ becomes infected is then (from (\ref{sis}) given 
by 
\[
1- (1-\rho q_A)^{\hat{D}_{A}}(1-\rho q_B)^{\hat{D}_{B}} (1- \rho  [(q_A  + q_B - f(2, \{A,B\})])^{\hat{D}_{AB}}.
\]
If the change is to network $\widehat{g}$ in which this individual is less multiplexed
then their connection profile is 
$(\hat{D}_{A}+a,\hat{D}_{B}+a,\hat{D}_{AB}-a)$ for some integer $a>0$, and then their probability
of being infected is
\[
1- (1-\rho q_A)^{\hat{D}_{A}+a}(1-\rho q_B)^{\hat{D}_{B}+a} (1- \rho  [(q_A  + q_B - f(2, \{A,B\})])^{\hat{D}_{AB}-a}.
\]
The second probability is larger than the first if and only if 
\[
(1-\rho q_A)^{a}(1-\rho q_B)^{a} (1- \rho  [(q_A  + q_B - f(2, \{A,B\})])^{-a}< 1,
\]
which simplifies to 
\[
(1-\rho q_A)(1-\rho q_B)< (1- \rho  [(q_A  + q_B - f(2, \{A,B\})]).
\]
This holds if and only if 
\[
\rho q_Aq_B < f(2, \{A,B\}),
\]
which is the claimed condition.\eproof

\bigskip

\noindent {\bf Proof of Proposition  \ref{prop:sis-simple}}:
Following the argument from the proof of Proposition \ref{prop:individualsimple}, 
for any $\rho$ equation \ref{sis-steady} has a higher solution for the less multiplexed type.   Thus, starting with the steady state $\rho$ for the more multiplexed distribution,  the new rates for all individuals are weakly and sometimes strictly higher for the less multiplexed distribution.  This leads to a higher $\rho'$.  Iterating, this converges upward for all types to a limit which is the steady state.  Conversely, if condition \ref{corr} is reversed, the convergence is downward for all types.\eproof
 
\bigskip

\noindent {\bf Proof of Proposition \ref{prop:individualcomplex}}:  It is enough to consider an individual $i$ with one change in their links where one layer of a multiplexed link to $j$ is reassigned to a neighbor $k$, so that $i$ is connected to $j$ on one layer and to $k$ on a different layer, where $i$ was initially not connected to $k$.
Our focus is on the pivotal cases:
\begin{enumerate}
    \item The number of infected messages $i$ has already received from other neighbors is either $\tau -1$ or $\tau - 2$.  (That both of these cases can occur with positive probability uses the condition that $\sum_{\ell,j} g_{ij}^\ell>\tau$, so that there are at least $\tau-1$ layer-connections from $i$ to others besides $j,k$.)
    \item At least one of the neighbors $j$ and $k$ is infected.
\end{enumerate}

The conditional probability (given that one is in one of these four cases) that $i$ gets infected can be found in the table below. The top entry in each cell represents the multiplexing scenario and the bottom represents the unmultiplexed case. 

\begin{table}[H]
\centering
\begin{tabular}{| cc | c | c |} \hline
& & $\mathrm{\tau - 1}$ & $\mathrm{\tau-2}$ \\ \hline
 &  &  &  \\
 &  & $q_A+q_B- f(2, \{A,B\}) $ & $f(2, \{A,B\})$ \\
 &  & $\downgeq$ & $\upgeq$ \\
& $\text{both $j,k$ infected}$ & $q_A+q_B- q_Aq_B$ & $q_Aq_B$ \\
&  &  &  \\ \cline { 2 - 4 }
 &  &  &  \\
& & $(q_A+q_B- f(2, \{A,B\}) )/2$ & $f(2, \{A,B\})/2$ \\
&  & $\downg$ & $\upg$ \\
& $\text{one of $j,k$ infected}$ & $(q_A+q_B)/2$ & $0$ \\
&  &  &  \\ \hline 
\end{tabular}
\end{table}

The inequality indicates which probability is larger.   The $\tau-1$ column (aggregating over both rows which have positive probability) has strictly higher probability for the unmultiplexed case, while the $\tau-2$ column has strictly higher probability for the multiplexed case.  
Let $\phi$ be the probability on the first column and $\psi$ on the second column, and note that the conditional probability of the first row is $\rho^2$ and the second row is $2\rho(1-\rho)$.
The differences in overall probabilities of infection of the multiplexed minus unmultiplexed is then
$$(\psi -\phi)\left[\rho^2 ( f(2, \{A,B\}) -q_Aq_B) + 
2\rho(1-\rho)f(2, \{A,B\}) /2 \right].
$$
Given that $f(2, \{A,B\}) -q_Aq_B \geq0$,
then this expression has the sign of $(\psi -\phi)$.
The proof is then completed by noting that for 
high enough $\rho, q_A, q_B$ the first column becomes more likely than the second,
and for low enough $\rho, q_A, q_B$ the second column becomes more likely than the first.This is where the condition that $(1+\varepsilon) q_A q_B \geq f(2,\{A,B\})\geq q_Aq_B$ is invoked.  With independent signal transmission across layers, for low enough $\rho, q_A, q_B$, it is strictly more probable to have fewer than more signals from the connections other than $j,k$, and thus $\psi-\phi>0$.  These probabilities are continuous in $f$ and so this holds for some $\varepsilon>0$.  The reverse is true for high enough $\rho, q_A, q_B$.\eproof

\bigskip

\noindent {\bf Proof of Proposition \ref{prop:SIScomplex}}:
We begin with the case of sufficiently low $q_A, q_B$ and high $\delta$.  In that case
$\rho$ will also be low (an absolute bound is simply $(q_A+q_B)/\delta$ as that is a crude upper bound on the infection rate of any given node that always has all neighbors infected and needs only one signal).   Then we can invoke Proposition \ref{prop:individualcomplex} for each connection configuration (noting that there are a finite number of them, taking the min over the $\underline{\rho}$), and then the remaining argument is analogous to the proof of Proposition  \ref{prop:sis-simple}.   The reverse holds for the case of sufficiently high $q_A, q_B$ and low $\delta$.\eproof

\begin{proposition}\label{prop:graph_acyclic}
The relation $\prec$ is acyclic.
\end{proposition}

\noindent {\bf Proof of Proposition \ref{prop:graph_acyclic}}
Recall that we denote the set of layers a link $ij$ belongs to by $\mathcal{L}_{ij}$. Define the total multiplexity index of a multigraph $g$ as $S_g = \sum_{i>j} |\mathcal{L}_{ij}|^2$.

We show that if $\widehat{g} \prec g$, then $S_g > S_{\widehat{g}}$.
By our definition of $\widehat{g} \prec g$, we know that there exist nodes $i,j,k$ and layers $\ell, \ell'$ such that ${g}_{ik}^\ell=0=
\widehat{g}_{ij}^\ell$ and $\widehat{g}_{ik}^\ell=1=
{g}_{ij}^\ell$, all else being equal. We only focus on the contribution of these edges in total multiplexing index since all other links are identical across the two multigraphs. For the multigraph $g$, this can be represented as $|\mathcal{L}_{ij}|^2 + |\mathcal{L}_{ik}|^2$, while for the less multiplexed graph $\widehat{g}$, the contribution of these edges can be written as $(|\mathcal{L}_{ij}|-1)^2 + (|\mathcal{L}_{ik}|+1)^2$. We can then write the difference in total multiplexing between $g$ and $\widehat{g}$ as 
\begin{align*}
    S_g - S_{\widehat{g}} &= |\mathcal{L}_{ij}|^2 + |\mathcal{L}_{ik}|^2 - (|\mathcal{L}_{ij}|-1)^2 - (|\mathcal{L}_{ik}|+1)^2 \\
    &= |\mathcal{L}_{ij}|^2 + |\mathcal{L}_{ik}|^2 - |\mathcal{L}_{ij}|^2 - |\mathcal{L}_{ik}|^2 - 2 + 2|\mathcal{L}_{ij}| - 2|\mathcal{L}_{ik}| \\
    &= 2(|\mathcal{L}_{ij}| - (|\mathcal{L}_{ik}| + 1))
\end{align*}
By  $\widehat{g} \prec g$, we know that $|\mathcal{L}_{ik}| < |\mathcal{L}_{ij}| - 1$ (recall that we assumed $i$ and $j$ were linked in at least two layers), hence $S_g >S_{\widehat{g}}$

Now, assume that there exists a cycle such that we have a sequence of multigraphs $g_i$ with $g_1 \prec g_2 \prec g_3 \prec \cdots \prec g_n \prec g_1$. But our proof implies $S_{g_1} < S_{g_2} < S_{g_3} < \cdots < S_{g_n} < S_{g_1}$, which gives us a contradiction. Hence the relation is acyclic.\eproof

\section{Supplementary Figures}

\begin{figure}
\centering
\includegraphics[scale = 0.7]{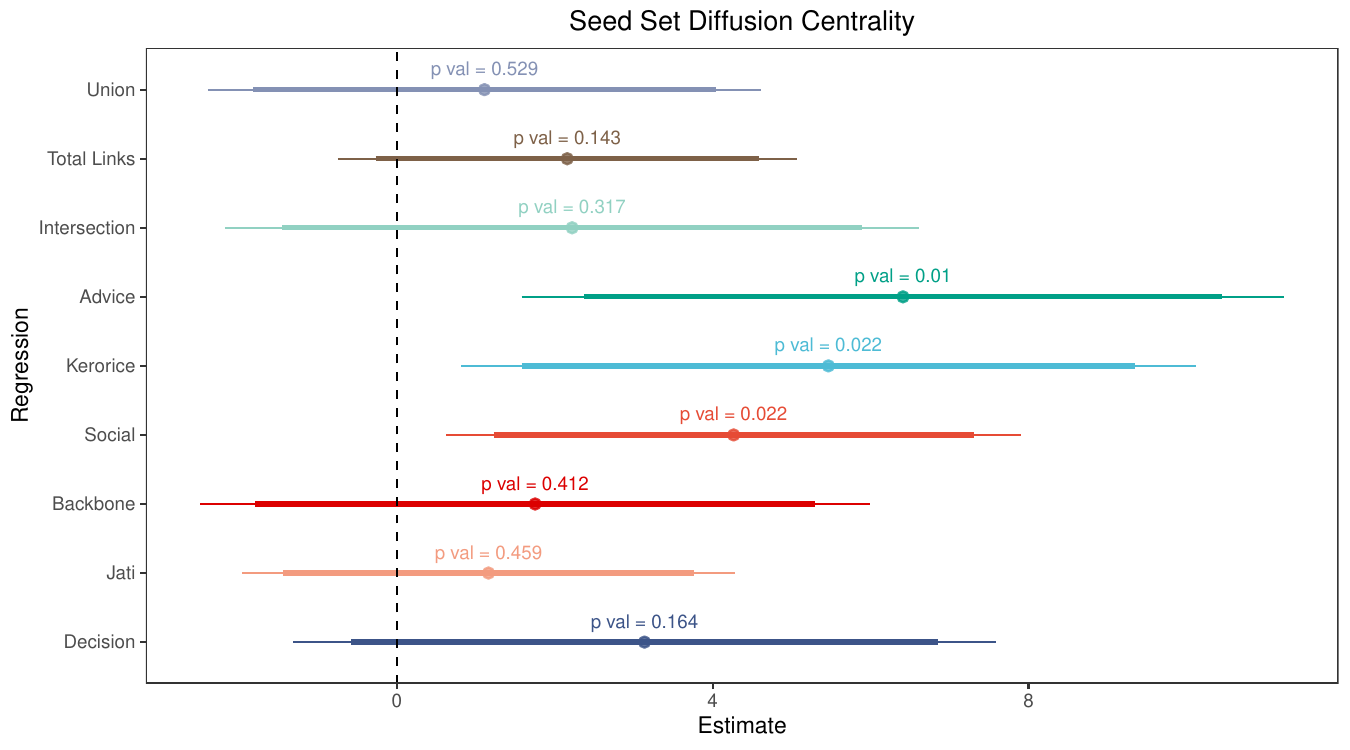}
\caption{
 OLS Estimates for Impact of Seed Set Diffusion Centrality}
\label{fig:ols}
\end{figure}

\

 Figure \ref{fig:ols} plots the results from Table \ref{tab:table_dcseed}.   $\hat{\beta}^\ell$ with both the 90\% and 95\% confidence intervals for each of the distinct layers. Seed centrality in the jati network is not statistically significantly associated with diffusion  ($p = 0.459$). Seed centrality in the advice, social, and kerorice networks all are significantly positively associated with diffusion.  The point estimates are large, roughly a 59\% increase.

\newpage

 \begin{figure}[h]
	\centering
	\subfloat[Scree Plot: Microfinance Villages]{%
		\includegraphics[width=0.5\linewidth]{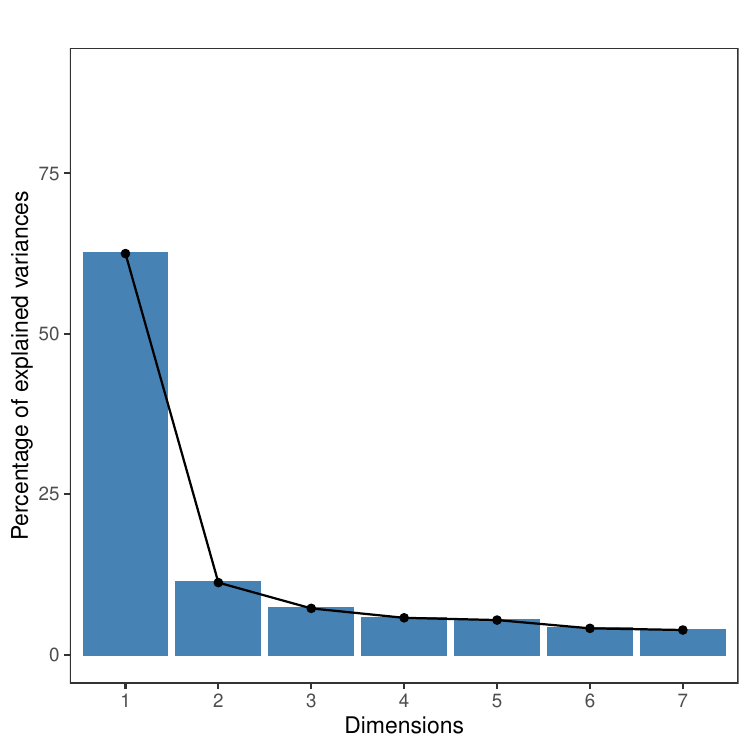}
	}
	
	\subfloat[Principal Components: Microfinance Villages]{%
		\includegraphics[width=0.5\linewidth]{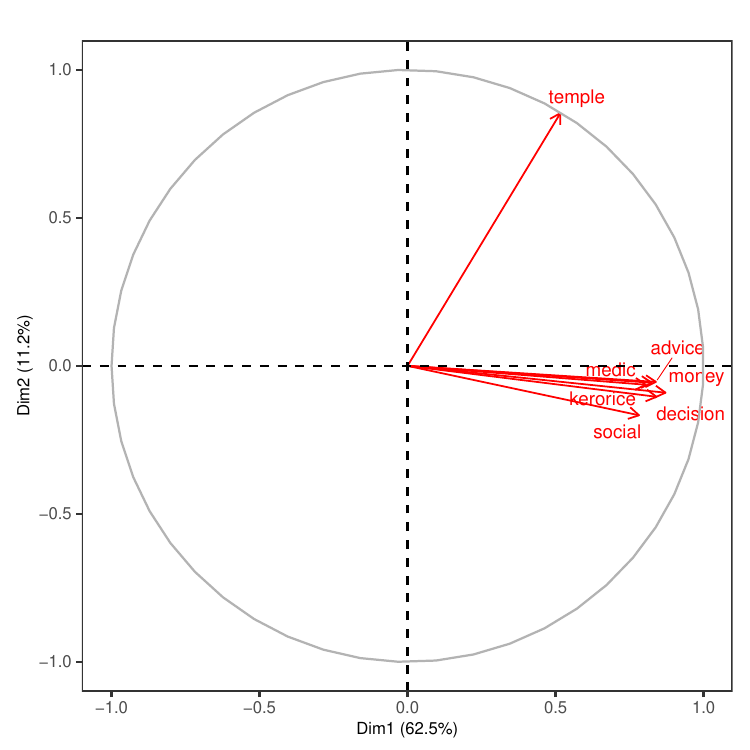}
	}
	
	\caption{Principal Component Analysis with the Temple Layer, but without Geography or Jati layers.}
	\label{fig:pca_withtemple}
\end{figure}

\newpage

\begin{figure}[H]
    \centering
    \includegraphics[scale = 0.8]{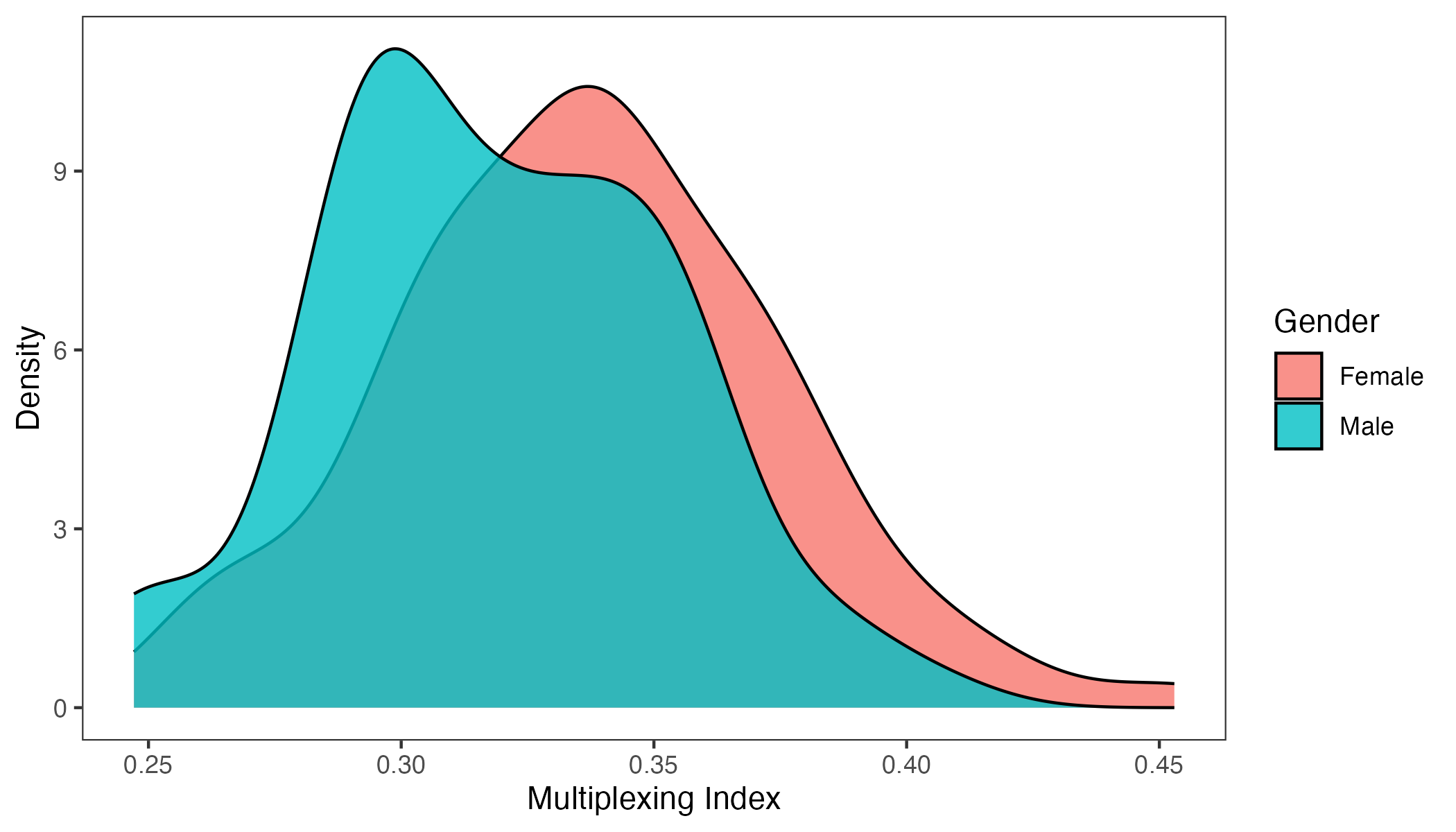}
    \caption{Multiplexing by gender.}
    \label{fig:mpex_gender_app}
\end{figure}

Here we redo the analysis from Figure \ref{fig:mpex_gender}, but instead we use the individual network data from Microfinance villages collected as part of Wave I of data collection in \cite{banerjeecdj2013} (instead of Wave II).\footnote{We have individual-level gender-distinguished data in the Wave I network survey, which elicited links from $46$\% of the households, giving us information on $70.84$\% of the links.}

\newpage
\subsection{Supplementary Tables}\label{supp:tables}

In Table \ref{tab:table_dcseed_nojati}, we redo Table \ref{tab:table_dcseed} but with the aggregate networks of union, intersection, and backbone constructed without including jati (the total link network is directed and never included jati).

\begin{table}[H]
\centering
\caption{Seed Set Diffusion Centrality (Jati excluded from aggregate layers)}
\label{tab:table_dcseed_nojati}
\scalebox{0.7}{
\begin{threeparttable}
\centering
\begin{tabular}[t]{lccccccccc}
\toprule
\multicolumn{1}{c}{ } & \multicolumn{9}{c}{No. Calls Received} \\
\cmidrule(l{3pt}r{3pt}){2-10}
  & 1 & 2 & 3 & 4 & 5 & 6 & 7 & 8 & 9\\
\midrule
Social & \num{4.266} &  &  &  &  &  &  &  & \\
 & (\num{1.820}) &  &  &  &  &  &  &  & \\
 & {}[\num{0.022}] &  &  &  &  &  &  &  & \\
Kero/Rice &  & \num{5.466} &  &  &  &  &  &  & \\
 &  & (\num{2.326}) &  &  &  &  &  &  & \\
 &  & {}[\num{0.022}] &  &  &  &  &  &  & \\
Advice &  &  & \num{6.410} &  &  &  &  &  & \\
 &  &  & (\num{2.416}) &  &  &  &  &  & \\
 &  &  & {}[\num{0.010}] &  &  &  &  &  & \\
Decision &  &  &  & \num{3.137} &  &  &  &  & \\
 &  &  &  & (\num{2.226}) &  &  &  &  & \\
 &  &  &  & {}[\num{0.164}] &  &  &  &  & \\
Jati &  &  &  &  & \num{1.161} &  &  &  & \\
 &  &  &  &  & (\num{1.559}) &  &  &  & \\
 &  &  &  &  & {}[\num{0.459}] &  &  &  & \\
Union &  &  &  &  &  & \num{2.868} &  &  & \\
 &  &  &  &  &  & (\num{1.994}) &  &  & \\
 &  &  &  &  &  & {}[\num{0.155}] &  &  & \\
Intersection &  &  &  &  &  &  & \num{4.492} &  & \\
 &  &  &  &  &  &  & (\num{1.996}) &  & \\
 &  &  &  &  &  &  & {}[\num{0.028}] &  & \\
Backbone &  &  &  &  &  &  &  & \num{5.851} & \\
 &  &  &  &  &  &  &  & (\num{2.575}) & \\
 &  &  &  &  &  &  &  & {}[\num{0.027}] & \\
Total Links &  &  &  &  &  &  &  &  & \num{2.158}\\
 &  &  &  &  &  &  &  &  & (\num{1.453})\\
 &  &  &  &  &  &  &  &  & {}[\num{0.143}]\\
\midrule
Num.Obs. & \num{68} & \num{68} & \num{68} & \num{68} & \num{68} & \num{68} & \num{68} & \num{68} & \num{68}\\
R2 & \num{0.194} & \num{0.254} & \num{0.313} & \num{0.161} & \num{0.110} & \num{0.145} & \num{0.227} & \num{0.263} & \num{0.131}\\
Dep Var mean & \num{8.691} & \num{8.691} & \num{8.691} & \num{8.691} & \num{8.691} & \num{8.691} & \num{8.691} & \num{8.691} & \num{8.691}\\
\bottomrule
\end{tabular}
\begin{tablenotes}
  \small
  \item \textit{Note:} Robust standard errors are given in parentheses and p-values in square brackets. Controls added: number of households, its powers, and a dummy for number of seeds in the village. Exogenous variables are the sum of Diffusion Centrality for seeds in each village for the layer. Exogenous variables have been standardized. None of the aggregate layers (union, intersection, backbone and total links) uses jati as an input.
  \end{tablenotes}
  \end{threeparttable}
}
\end{table}

\newpage

\begin{table}
\centering
\caption{Component Loadings: Microfinance Villages}
\begin{tabular}[t]{llllllllll}
\toprule
Network & PC1 & PC2 & PC3 & PC4 & PC5 & PC6 & PC7 & PC8 & PC9\\
\midrule
\midrule
social & 0.37 & -0.04 & -0.03 & 0.18 & -0.55 & 0.69 & -0.13 & -0.17 & -0.07\\
kerorice & 0.38 & 0.02 & 0.01 & 0.07 & -0.43 & -0.69 & -0.38 & -0.17 & -0.11\\
money & 0.41 & 0.06 & 0.02 & 0.11 & 0.07 & 0.01 & -0.14 & 0.72 & 0.52\\
advice & 0.40 & 0.08 & 0.03 & 0.07 & 0.51 & 0.11 & -0.22 & 0.16 & -0.70\\
decision & 0.40 & 0.07 & 0.02 & 0.12 & 0.47 & 0.03 & -0.01 & -0.62 & 0.45\\
medic & 0.39 & 0.05 & 0.01 & 0.07 & -0.12 & -0.17 & 0.88 & 0.04 & -0.14\\
temple & 0.24 & 0.06 & 0.02 & -0.96 & -0.05 & 0.08 & -0.02 & -0.03 & 0.03\\
jati & 0.10 & -0.66 & -0.74 & -0.04 & 0.08 & -0.04 & 0.01 & 0.02 & 0.00\\
distance & -0.07 & 0.73 & -0.68 & 0.02 & -0.05 & 0.01 & -0.02 & -0.01 & 0.00\\
\bottomrule
\end{tabular}
\label{tab:loadings_mf}
\end{table}

\begin{table}
\centering
\caption{Component Loadings: RCT Villages}
\begin{tabular}[t]{llllll}
\toprule
Network & PC1 & PC2 & PC3 & PC4 & PC5\\
\midrule
\midrule
social & 0.49 & 0.03 & -0.58 & 0.52 & -0.39\\
kerorice & 0.50 & 0.04 & -0.39 & -0.48 & 0.60\\
advice & 0.50 & 0.05 & 0.37 & -0.51 & -0.59\\
decision & 0.49 & 0.05 & 0.61 & 0.50 & 0.37\\
jati & 0.09 & -1.00 & 0.02 & 0.00 & 0.00\\
\bottomrule
\end{tabular}
\label{tab:loadings_rfe}
\end{table}


\newpage

\begin{table}[H]

\caption{F-test for the layers}
\label{tab:ftest_short}
\centering
\begin{threeparttable}
\begin{tabular}[t]{lrrrrrr}
\toprule
layer & df & R.sq. & F-stat & p-val & F-stat marginal & p-val marginal\\
\midrule
Advice & 1 & 0.233 & 20.057 & 0.000 &  & \\
Jati & 2 & 0.263 & 2.628 & 0.110 & 2.628 & 0.110\\
Decision & 3 & 0.272 & 1.728 & 0.186 & 0.834 & 0.365\\
Kero/Rice & 4 & 0.293 & 1.768 & 0.162 & 1.804 & 0.184\\
Social & 5 & 0.293 & 1.306 & 0.278 & 0.006 & 0.938\\
\bottomrule
\end{tabular}
\begin{tablenotes}
\small
\item \textit{Note}: The ``F-stat'' and ``p-val'' columns correspond to the cumulative test comparing each specification with the intercept only benchmark. ``F-stat marginal'' and ``p-val marginal'' columns correspond to the marginal test when adding a given layer. Relative to Table \ref{tab:ftest1} here we exclude the ``constructed" intersection, union, total links, and backbone layers.
\end{tablenotes}
\end{threeparttable}
\end{table}

\subsection{Algorithms} 

\begin{algorithm}[htbp]
\SetAlgoLined
\KwIn{Multiplexed network’s adjacency matrix \( G = \{G^{(1)}, G^{(2)} \} \), transmission probability \( q \), infection threshold \( \tau \), recovery probability \( \delta \), initial set of infected nodes \( I_0 \)}
\KwOut{Share of infected nodes in steady state}
\BlankLine

\textbf{Definitions:}
\begin{itemize}
    \item \( N \): Set of all nodes in the network, \( |N| = n \)
    \item \( S_t \): Set of susceptible nodes at time \( t \)
    \item \( I_t \): Set of infected nodes at time \( t \)
    \item \( \sigma_{i,t} \): State of node \( i \) at time \( t \), where \( \sigma_{i,t} \in \{S, I\} \)
    \item \( E_{i,t} \): Number of exposures (infections) node \( i \) is exposed to at time \( t \)
\end{itemize}
\BlankLine

\textbf{Step 1:} Initialize \( S_0 = N \setminus I_0 \), \( I_0 \)\;
\textbf{Step 2:} \While{t $< $ 1000}{
    \ForEach{\( i \in S_t \)}{
        Calculate \( E_{i,t} = \sum_{j \in N} \sum_{l=1}^2 G_{ij}^{(l)} \cdot \mathbb{I}(\sigma_{j,t} = I) \cdot B_{ij,t}^{(l)} \), where \( B_{ij,t}^{(l)} \sim \text{Bernoulli}(q) \) i.i.d.\;
        \If{\( E_{i,t} \geq \tau \)}{
            Node \( i \) becomes infected: \( \sigma_{i,t+1} = I \)\;
        }
    }
    \ForEach{\( i \in I_t \)}{
        Node \( i \) recovers with probability \( \delta \): \( \sigma_{i,t+1} = S \) with probability \( \delta \)\;
    }
    Update \( S_{t+1} = \{i \in N \mid \sigma_{i,t+1} = S\} \)\;
    Update \( I_{t+1} = \{i \in N \mid \sigma_{i,t+1} = I\} \)\;
    \If{$abs(\frac{|I_{t+1}|}{n} - \frac{|I_{t}|}{n}) < 1e-8$}{
            \textbf{break}\;}
}
\textbf{Step 4:} After convergence, run the simulation for an additional 100 iterations to stabilize the results and take the average across these iterations\;
\caption{Diffusion Simulation on Multiplexed Networks}  \label{alg: graph_diffusion}
\end{algorithm}

\clearpage

\begin{algorithm}[H]
\SetAlgoLined
\KwIn{Three network layers represented as adjacency matrices: \( A_1 \), \( A_2 \), \( A_3 \)}
\KwOut{Two multiplexed networks \( M_1 \) and \( M_2 \)}
\BlankLine
\textbf{Step 1:} Use directed \texttt{Kerorice}, \texttt{social}, and \texttt{advice} as the three matrices respectively\;
\textbf{Step 2:} Sort \( A_2 \) and \( A_3 \) based on their average out-degree, in descending order\;
\textbf{Step 3:} Prune the network with the higher average out-degree (among \( A_2 \) and \( A_3 \)) to match that of the network with the lower average out-degree. Denote the pruned network as \( A_2' \)\;
\textbf{Step 4:} Generate the first multiplexed network, \( M_1 \), by combining the adjacency matrices of \( A_1 \) and the unpruned network (either \( A_2 \) or \( A_3 \), whichever had the lower out-degree)\;
\textbf{Step 5:} Generate the second multiplexed network, \( M_2 \), by combining the adjacency matrices of \( A_1 \) and \( A_2' \)\;
\caption{Multiplexed Network Generation} \label{alg: graph_rewire}
\end{algorithm}

\end{document}